\documentclass[11pt]{article}
\def\preprintversion{}
\newcommand{\preprintauthors}{%
  Christian Pipal\textsuperscript{*}, Morgan Wack, and Eva-Maria Vogel\\[0.5em]
  {\small Department of Communication and Media Research (IKMZ)}\\
  {\small University of Zurich, Switzerland}\\[0.5em]
  {\small\textsuperscript{*}Corresponding author: \href{mailto:c.pipal@ikmz.uzh.ch}{c.pipal@ikmz.uzh.ch}}%
}
\AtBeginDocument{%
  \hypersetup{pdfauthor={Christian Pipal, Morgan Wack, Eva-Maria Vogel},
    pdftitle={Who Gets Counted? Large Language Models, Informative Refusal, and Downstream Bias in Name-Based Demographic Classification}}%
  \addtocontents{toc}{\protect\setcounter{tocdepth}{-1}}%
}
\newcommand{\preprintappendix}{%
  \clearpage
  \singlespacing
  \appendix
  \renewcommand{\thetable}{\Alph{section}.\arabic{table}}%
  \renewcommand{\thefigure}{\Alph{section}.\arabic{figure}}%
  \renewcommand{\theHtable}{appendix.\Alph{section}.\arabic{table}}%
  \renewcommand{\theHfigure}{appendix.\Alph{section}.\arabic{figure}}%
  \counterwithin*{table}{section}%
  \counterwithin*{figure}{section}%
  \addtocontents{toc}{\protect\setcounter{tocdepth}{2}}%
  \let\preprintcontents\tableofcontents
  \renewcommand{\tableofcontents}{\begingroup\small\preprintcontents\endgroup}%
  \begin{center}
{\LARGE\textbf{Supplementary Information}}

\vspace{1em}

{\large Who Gets Counted? Large Language Models, Informative Refusal, and\\Downstream Bias in Name-Based Demographic Classification}
\end{center}

\vspace{1em}

\tableofcontents
\newpage

\section{LLM Prompts and Prompt Sensitivity}
\label{app:prompts}

The hosted models received a single user message consisting of the instruction, a blank line, and the input. The input was \texttt{Author: [name]} followed by \texttt{Institution Country: [code]}, with the code \texttt{US} for voters, \texttt{CN} for Study~1, and the author's country of affiliation for Study~3. The local model received the same instruction as a system message and the input as the user message. The race instruction was

\begin{quote}\small
You are classifying the likely race/ethnicity of academic authors based on their names and institutional affiliations. For each author, provide your best estimate from the following categories: White, Black, Hispanic, Asian, Middle Eastern, South Asian, or Unknown. Respond with only the category name.
\end{quote}

\noindent and the gender instruction was

\begin{quote}\small
You are classifying the likely gender of academic authors based on their names. For each author, provide your best estimate: Male, Female, or Unknown. Respond with only: Male, Female, or Unknown.
\end{quote}

\noindent The instructions did not request a confidence level. We chose these baseline instructions (v1) from four variants tested on 200 names per task. The race variants were (v2) an academic system message stating that classifications are aggregated for research and followed by a shortened request, (v3) a request for ``ethnic or cultural background'' with the categories European-descent, African-descent, Hispanic/Latino, East Asian, South Asian, Middle Eastern, or Uncertain, and (v4) the baseline categories preceded by five labeled example names. The gender variants were (v2) a system message describing research on Chinese scholarly publishing, (v3) a question about which gender a Pinyin name is more commonly associated with, and (v4) five labeled example Pinyin names. The full text of every variant is in the replication code. Tables~\ref{tab:prompt_race} and~\ref{tab:prompt_gender} report coverage and conditional accuracy. Every variant that raised coverage lowered conditional accuracy.

\begin{table}[htbp]
\centering
\caption{Prompt sensitivity, race classification (200 North Carolina voters)}
\label{tab:prompt_race}
\small
\begin{tabular}{lcccc}
\toprule
& \multicolumn{2}{c}{Claude Haiku 4.5} & \multicolumn{2}{c}{Gemini 2.5 Flash} \\
\cmidrule(lr){2-3} \cmidrule(lr){4-5}
Variant & Coverage & Cond.\ acc. & Coverage & Cond.\ acc. \\
\midrule
v1 Baseline & 50\% & 80\% & 93\% & 69\% \\
v2 Academic framing & 82\% & 61\% & 99\% & 66\% \\
v3 Alternative terminology & 1\% & --- & 90\% & 67\% \\
v4 Few-shot examples & 82\% & 59\% & 95\% & 67\% \\
\bottomrule
\end{tabular}
\end{table}

\begin{table}[htbp]
\centering
\caption{Prompt sensitivity, gender classification (200 Chinese names)}
\label{tab:prompt_gender}
\small
\begin{tabular}{lcccc}
\toprule
& \multicolumn{2}{c}{Claude Haiku 4.5} & \multicolumn{2}{c}{Gemini 2.5 Flash} \\
\cmidrule(lr){2-3} \cmidrule(lr){4-5}
Variant & Coverage & Cond.\ acc. & Coverage & Cond.\ acc. \\
\midrule
v1 Baseline & 8\% & 94\% & 26\% & 94\% \\
v2 Chinese academic framing & 67\% & 75\% & 62\% & 86\% \\
v3 Majority-gender framing & 100\% & 55\% & 81\% & 81\% \\
v4 Few-shot examples & 49\% & 77\% & 43\% & 92\% \\
\bottomrule
\end{tabular}
\end{table}

\newpage
\section{Model Versions, Settings, and Query Dates}
\label{app:models}

\begin{table}[htbp]
\centering
\caption{LLM identifiers, settings, and query dates}
\label{tab:models}
\footnotesize
\begin{tabularx}{\textwidth}{>{\raggedright\arraybackslash}p{2.3cm}>{\raggedright\arraybackslash}p{3.4cm}>{\raggedright\arraybackslash}X>{\raggedright\arraybackslash}p{2.2cm}}
\toprule
Model & Identifier & Access and settings & Queried \\
\midrule
Claude Haiku 4.5 & \texttt{claude-haiku-4-5-}\newline\texttt{20251001} & Anthropic Messages API, temperature 0, maximum 50 output tokens, no extended thinking. Training data cutoff July 2025. & 9 and 10 March 2026 \\
\addlinespace
Gemini 2.5 Flash & \texttt{gemini-2.5-flash} & Google Generative Language API (v1beta), temperature 0, maximum 256 output tokens including thinking tokens, default thinking budget. Knowledge cutoff January 2025. & 9 and 10 March 2026 \\
\addlinespace
Qwen 2.5 7B Instruct & \texttt{qwen2.5:7b-instruct-}\newline\texttt{q4\_K\_M} & Ollama 0.20.5 on a local laptop (Apple M5 Pro, 48 GB), temperature 0, seed 42, maximum 12 output tokens. Released September 2024. & 5 August 2026 \\
\bottomrule
\end{tabularx}
\end{table}

\noindent Training cutoffs are from the providers' documentation (Anthropic's model overview and Google's Gemini 2.5 Flash model card). Query dates are the dates on which the prediction files were written.

\newpage
\section{Determinism}
\label{app:determinism}

We re-classified 200 Study~2 voters (50 per group) with both hosted models at temperature zero. Of the 95 names Haiku labeled in both runs, all 95 received the same label, and 101 of its 103 original refusals recurred. Of the 185 names Gemini labeled in both runs, 184 received the same label. Temperature zero implements greedy decoding, but floating-point nondeterminism on shared hardware means identical output is not guaranteed for hosted models. For the local model we re-ran the first 2,000 Study~2 voters, and every answer, including the unlabeled ones, was identical.

\newpage
\section{Why LLM Labels Are Missing}
\label{app:missing}

A name can lack an LLM label for three reasons. The model can refuse, returning an empty answer or a statement that it will not classify people by race. It can answer ``Unknown'', the abstention option the instruction offers. Or it can answer with a category that the four-category scheme cannot use, mainly Middle Eastern. Table~\ref{tab:missing_reasons} separates the three. For race, a fifth of Haiku's answers on voters are explicit refusals. For gender, the models almost never refuse, and nearly all missing labels are ``Unknown''.

\begin{table}[!htbp]
\centering
\caption{Why LLM labels are missing, share of all names by model and study. An explicit refusal is an empty answer or a statement that the model will not classify. ``Unknown'' is the abstention option offered in the prompt. Outside categories are answers the four-category scheme cannot use (for race, mainly Middle Eastern).}
\label{tab:missing_reasons}
\small
\begin{tabular}{llcccc}
\toprule
Study & Model & Labeled & Explicit refusal & ``Unknown'' & Outside categories \\
\midrule
Study 1 (gender) & Claude Haiku 4.5 & 8.2 & 0.0 & 91.8 & 0.0 \\
Study 1 (gender) & Gemini 2.5 Flash & 28.4 & 0.0 & 71.6 & 0.0 \\
Study 1 (gender) & Qwen 2.5 7B & 7.6 & 0.0 & 92.3 & 0.0 \\
Study 2 (race) & Claude Haiku 4.5 & 49.0 & 20.9 & 29.0 & 1.1 \\
Study 2 (race) & Gemini 2.5 Flash & 89.6 & 0.0 & 8.9 & 1.5 \\
Study 2 (race) & Qwen 2.5 7B & 94.8 & 0.2 & 2.5 & 2.5 \\
Study 3 (race) & Claude Haiku 4.5 & 66.0 & 10.3 & 15.7 & 8.0 \\
Study 3 (race) & Gemini 2.5 Flash & 86.3 & 0.0 & 5.3 & 8.3 \\
Study 3 (race) & Qwen 2.5 7B & 75.0 & 1.0 & 10.7 & 13.3 \\
Study 3 (gender) & Claude Haiku 4.5 & 95.7 & 0.0 & 4.3 & 0.0 \\
Study 3 (gender) & Gemini 2.5 Flash & 90.0 & 0.0 & 10.0 & 0.0 \\
Study 3 (gender) & Qwen 2.5 7B & 82.3 & 0.0 & 17.7 & 0.0 \\
\bottomrule
\end{tabular}
\end{table}

The main analysis maps East, South, and Southeast Asian answers to Asian and treats Middle Eastern answers as no label. The U.S.\ Census counts Middle Eastern and North African origin as White. Mapping Middle Eastern answers to White raises coverage by at most 1.9 points and changes conditional accuracy by at most 1.1 points (Table~\ref{tab:mapping}). It changes the bias in the registration gap by less than 0.01 and lowers the bias in the age gap, from 6.0 to 5.4 years for Haiku and from 0.5 to 0.0 years for Qwen.

\begin{table}[!htbp]
\centering
\caption{Sensitivity to the mapping of Middle Eastern answers, Study 2. The main analysis treats them as no label. The alternative maps them to White, following the U.S.\ Census convention for Middle Eastern and North African origin. Coverage and conditional accuracy are on the four-category sample, and the biases are for hard labels.}
\label{tab:mapping}
\small
\begin{tabular}{llcccc}
\toprule
Model & Middle Eastern & Coverage & Cond.\ accuracy & Bias($\hat{\Delta}_{\text{party}}$) & Bias($\hat{\Delta}_{\text{age}}$) \\
\midrule
Claude Haiku 4.5 & No label & 50.5 & 89.3 & $+0.191$ & $+6.04$ \\
Claude Haiku 4.5 & White & 51.1 & 88.4 & $+0.197$ & $+5.37$ \\
Gemini 2.5 Flash & No label & 90.7 & 78.6 & $+0.271$ & $+1.42$ \\
Gemini 2.5 Flash & White & 91.6 & 78.0 & $+0.272$ & $+1.10$ \\
Qwen 2.5 7B & No label & 95.6 & 73.5 & $+0.344$ & $+0.53$ \\
Qwen 2.5 7B & White & 97.5 & 72.4 & $+0.344$ & $+0.04$ \\
\bottomrule
\end{tabular}
\end{table}

\newpage
\section{LLM Name-Component Ablation}
\label{app:ablation}

To separate what the LLMs know from what they are told, we classified 2,000 Study~2 voters (500 per group) under four inputs. All four conditions supply the country code \texttt{US}. The last condition adds the voter's county of residence to the name (``JAMES SMITH from MECKLENBURG county, NC'').

\begin{table}[htbp]
\centering
\caption{Coverage and conditional accuracy by input (2,000 Study 2 voters)}
\label{tab:ablation_overall}
\footnotesize
\begin{tabular}{lcccc}
\toprule
Input & Haiku coverage & Haiku cond.\ acc. & Gemini coverage & Gemini cond.\ acc. \\
\midrule
Surname only & 28.8\% & 92.9\% & 70.3\% & 79.4\% \\
First name only & 9.9\% & 96.0\% & 56.7\% & 70.2\% \\
Full name & 53.8\% & 89.7\% & 91.7\% & 79.7\% \\
Full name and county of residence & 31.1\% & 94.1\% & 93.1\% & 81.0\% \\
\bottomrule
\end{tabular}
\end{table}

Haiku labels more names as the name gets more informative, from 9.9\% with first names to 53.8\% with full names, while its conditional accuracy stays near 90\%. Adding the county of residence has the opposite effect. Haiku's coverage falls to 31.1\%, although Gemini's accuracy shows that the county carries useful information. A request that looks more like a query about an identifiable individual draws more refusals, which is consistent with a role for safety training, though the ablation does not isolate that mechanism. Surnames carry more signal than first names for every group, but first names add signal for Black voters. Among the Black voters it labels, Gemini is correct on 43.1\% from first names and on 31.7\% from surnames.

\newpage
\section{Tool Inputs and Category Mapping}
\label{app:tools}

\begin{table}[htbp]
\centering
\caption{Tool inputs and mapping to the four categories}
\label{tab:crosswalk}
\footnotesize
\begin{tabularx}{\textwidth}{>{\raggedright\arraybackslash}p{3cm}>{\raggedright\arraybackslash}p{3.8cm}>{\raggedright\arraybackslash}X}
\toprule
Tool & Input & Mapping \\
\midrule
WRU (surname) & Surname & Most likely of five Census categories. ``Other'' counts as no label. \\
WRU (+county) & Surname and county & As above, with county priors from the 2020 Census. \\
WRU (+tract) & Surname and census tract & As above, with tract priors for the 79\% of voters whose address geocoded to a tract and county priors for the rest. \\
predictrace & Surname & Census surname lookup. Surnames not in the list receive no label. \\
rethnicity & First name and surname & Four categories returned directly. \\
ethnicolr & First name and surname & Florida model. Asian subcategories collapsed to Asian. \\
pyethnicity & First name and surname & Most likely of four categories. \\
LLMs & Full name and country & East, South, and Southeast Asian collapsed to Asian. Middle Eastern, ``Unknown'', and refusals count as no label. \\
\bottomrule
\end{tabularx}
\end{table}

\noindent Accuracy in Table~\ref{tab:study2_overall_v2} is computed on the 17,998 voters who report White, Black, Hispanic, or Asian identity. The 2,000 voters in the residual category are analyzed in SI~\ref{app:other}. The downstream estimands in Section~\ref{sec:error_propagation} are defined on all 19,998 sampled voters.

\newpage
\section{Coverage Definitions}
\label{app:coverage}

Coverage means different things for different tools. For the Census-based tools a name goes unlabeled when its surname is missing from the Census surname list or when the most likely category is ``Other''. The LSTM tools label every name by construction. The LLMs leave names unlabeled when they refuse, answer ``Unknown'', or answer outside the four categories (SI~\ref{app:missing}). Haiku leaves about half of the voters unlabeled for race and most Chinese names unlabeled for gender, while it labels almost all Study~3 authors for gender. Conditional accuracy is always computed among labeled names, and total accuracy counts unlabeled names as errors.

\newpage
\section{Scoring Targets in Study 1}
\label{app:ceiling}

Study~1 knows the gender composition of each name's bearers but not the gender of any individual. Two scoring targets follow, and they answer different questions. Majority agreement asks whether a tool returns the majority gender of a name. A perfect classifier could agree with the majority for every name, so this target has no ceiling below 100\%. Individual accuracy asks how often a tool's label would be right for a person drawn from the bearers of a labeled name, and a female label earns the name's female share. No name-only classifier can do better on average than always picking the majority gender, so the mean majority share of the labeled names is the ceiling for this target. Because tools label different names, each tool has its own ceiling.

Both targets can be averaged over names, with each sampled name counted once, or over people, with each name weighted by its number of bearers in the source data. Name weighting describes the stratified sample of names. Person weighting gives more weight to common names, which are less ambiguous on average. Table~\ref{tab:study1_weighting} reports both. Across all 10,000 names the individual-level ceiling is 79.7\% weighted by names and 80.6\% weighted by people. No tool reaches its ceiling for individual accuracy. Genderize.io comes within 14.2 points and the LLMs within about 5 to 9 points, and on the names both label, Gemini and Genderize.io differ by 4.5 points in individual accuracy (78.3\% against 73.8\%).

\begin{table}[htbp]
\centering
\caption{Study 1 scores weighted by names and by people. Name-weighted figures count each sampled name once. Person-weighted figures weight each name by its number of bearers in the source data. Agreement is with the majority gender, individual accuracy is the expected accuracy for a bearer of a labeled name, and the ceiling is the mean majority share of the labeled names.}
\label{tab:study1_weighting}
\small
\begin{tabular}{lcccccc}
\toprule
 & \multicolumn{3}{c}{Name-weighted} & \multicolumn{3}{c}{Person-weighted} \\
\cmidrule(lr){2-4}\cmidrule(lr){5-7}
Tool & Agreement & Individual & Ceiling & Agreement & Individual & Ceiling \\
\midrule
Genderize.io & 70.4 & 64.8 & 79.0 & 84.2 & 74.1 & 80.5 \\
\texttt{gender} R package & 56.5 & 55.7 & 81.0 & 77.8 & 65.1 & 76.7 \\
\midrule
Gemini 2.5 Flash & 87.7 & 78.7 & 84.0 & 97.6 & 86.8 & 87.7 \\
Claude Haiku 4.5 & 85.1 & 78.9 & 85.6 & 91.6 & 84.3 & 89.3 \\
Qwen 2.5 7B (local) & 81.7 & 75.3 & 84.2 & 85.5 & 79.8 & 89.6 \\
\bottomrule
\end{tabular}
\end{table}

\newpage
\section{Genderize.io Country Parameter}
\label{app:genderize}

We queried Genderize.io without its optional \texttt{country\_id} parameter. Re-running all 10,000 names with \texttt{country\_id=CN} raises conditional accuracy from 70.4\% to 73.7\% and lowers coverage from 79.3\% to 50.0\% (Table~\ref{tab:genderize_cn}). Recall for female names rises from 75.8\% to 80.6\% among labeled names, which suggests that the global model leans male for Chinese names. We use the global configuration because it labels far more names.

\begin{table}[htbp]
\centering
\caption{Genderize.io with and without country localization (Study 1)}
\label{tab:genderize_cn}
\small
\begin{tabular}{lcc}
\toprule
Configuration & Coverage & Conditional accuracy \\
\midrule
Without \texttt{country\_id} (used in the paper) & 79.3\% & 70.4\% \\
With \texttt{country\_id=CN} & 50.0\% & 73.7\% \\
\bottomrule
\end{tabular}
\end{table}

\newpage
\section{Study 1 by Name Ambiguity}
\label{app:study1ambiguity}

Table~\ref{tab:study1_ambiguity} stratifies Study~1 by the majority-gender share of each name. In the most ambiguous stratum every tool's individual accuracy lies between 50.9\% and 52.0\%, a few points below the ceiling of about 55\%, so no tool knows much about these names. The LLMs' advantage lies in the unambiguous stratum, where Haiku and Gemini agree with the majority gender on 94.4\% and 95.7\% of the names they label and come within 5 and 4 points of the individual-level ceiling.

\begin{table}[!htbp]
\centering
\caption{Study 1 performance stratified by name ambiguity. Strata are defined by the majority-gender share in the source data. Coverage is the share of names receiving a label. Majority agreement is the share of labeled names whose label is the name's majority gender. Individual accuracy is the expected accuracy for a person drawn from the bearers of a labeled name (a female label earns the name's female share). The individual-level ceiling is the mean majority share, for the whole stratum in the panel heading and for the tool's labeled names in the last column. All quantities are averaged over names.}
\label{tab:study1_ambiguity}
\small
\begin{tabular}{lcccc}
\toprule
 & & Majority & Individual & Ceiling on \\
Tool & Coverage & agreement & accuracy & labeled names \\
\midrule
\multicolumn{5}{l}{\textit{Majority share 50--60\%} ($N{=}$1,496, individual-level ceiling 55.5\%)} \\
Genderize.io & 83.4 & 58.1 & 50.9 & 55.3 \\
gender R package & 3.9 & 51.7 & 51.1 & 55.6 \\
Claude Haiku 4.5 & 4.8 & 56.9 & 51.4 & 55.2 \\
Gemini 2.5 Flash & 19.2 & 69.0 & 52.0 & 55.5 \\
Qwen 2.5 7B (local) & 4.9 & 59.5 & 51.1 & 55.4 \\
\addlinespace
\multicolumn{5}{l}{\textit{Majority share 60--80\%} ($N{=}$3,504, individual-level ceiling 71.1\%)} \\
Genderize.io & 81.4 & 62.2 & 55.4 & 70.9 \\
gender R package & 3.1 & 48.2 & 48.3 & 70.7 \\
Claude Haiku 4.5 & 5.0 & 66.1 & 57.7 & 71.0 \\
Gemini 2.5 Flash & 20.4 & 74.5 & 60.8 & 71.3 \\
Qwen 2.5 7B (local) & 5.5 & 66.5 & 56.7 & 70.6 \\
\addlinespace
\multicolumn{5}{l}{\textit{Majority share >80\%} ($N{=}$5,000, individual-level ceiling 93.0\%)} \\
Genderize.io & 76.6 & 80.5 & 76.3 & 92.7 \\
gender R package & 4.1 & 62.3 & 60.9 & 93.5 \\
Claude Haiku 4.5 & 11.5 & 94.4 & 88.8 & 93.8 \\
Gemini 2.5 Flash & 36.8 & 95.7 & 89.8 & 93.4 \\
Qwen 2.5 7B (local) & 9.9 & 90.9 & 86.2 & 93.8 \\
\addlinespace
\bottomrule
\end{tabular}
\end{table}

\newpage
\section{Per-Group Accuracy in Study 2}
\label{app:pergroup}

Table~\ref{tab:study2_race} reports conditional accuracy within each self-reported group, which scores each tool on the voters it labels. Table~\ref{tab:study2_total} reports coverage and total accuracy by group, which count unlabeled voters as errors. The two views differ most for Haiku. It is correct on 67.2\% of the Black voters it labels but labels only 25.0\% of Black voters, so it labels 16.8\% of all Black voters correctly. Its Black precision of 95.8\% is the highest of any tool, and its Black F1 falls from 0.79 among labeled voters to 0.29 when unlabeled Black voters count as missed. For tools that label nearly every voter the two versions of F1 are almost identical.

\begin{table}[!htbp]
\centering
\caption{Per-group conditional accuracy (percent) with 95\% bootstrap intervals and Black F1, Study 2 four-category sample ($N = 17{,}998$). Each accuracy cell is the share of labeled voters in the group who receive the correct label. Black F1 combines precision and recall for the Black category. The labeled version computes recall among the Black voters the tool labels. The full-sample version counts unlabeled Black voters as missed, so it reflects coverage as well as accuracy.}
\label{tab:study2_race}
\footnotesize
\begin{tabular}{lcccccc}
\toprule
 & & & & & \multicolumn{2}{c}{Black F1} \\
\cmidrule(lr){6-7}
Tool & White & Black & Hispanic & Asian & Labeled & All \\
\midrule
pyethnicity & 87.7 {\tiny[86.7, 88.6]} & 64.1 {\tiny[62.7, 65.5]} & 84.8 {\tiny[83.8, 85.8]} & 76.5 {\tiny[75.0, 78.0]} & 0.72 & 0.72 \\
WRU (+tract) & 88.1 {\tiny[87.2, 89.0]} & 62.8 {\tiny[61.4, 64.1]} & 78.3 {\tiny[77.1, 79.4]} & 62.6 {\tiny[60.8, 64.3]} & 0.69 & 0.69 \\
rethnicity & 66.3 {\tiny[65.0, 67.6]} & 71.5 {\tiny[70.2, 72.7]} & 79.2 {\tiny[78.1, 80.4]} & 80.5 {\tiny[79.1, 82.0]} & 0.69 & 0.69 \\
WRU (+county) & 85.5 {\tiny[84.5, 86.5]} & 54.7 {\tiny[53.4, 56.2]} & 78.4 {\tiny[77.2, 79.4]} & 59.9 {\tiny[58.1, 61.6]} & 0.62 & 0.61 \\
ethnicolr & 95.1 {\tiny[94.4, 95.6]} & 37.0 {\tiny[35.7, 38.4]} & 81.3 {\tiny[80.3, 82.3]} & 51.4 {\tiny[49.7, 53.2]} & 0.51 & 0.51 \\
WRU (surname) & 95.7 {\tiny[95.1, 96.2]} & 16.7 {\tiny[15.6, 17.7]} & 79.2 {\tiny[78.1, 80.4]} & 61.5 {\tiny[59.7, 63.1]} & 0.28 & 0.28 \\
predictrace & 96.1 {\tiny[95.5, 96.6]} & 16.1 {\tiny[15.1, 17.2]} & 77.1 {\tiny[75.7, 78.5]} & 73.4 {\tiny[71.6, 75.2]} & 0.27 & 0.26 \\
\midrule
Gemini 2.5 Flash & 90.9 {\tiny[90.1, 91.8]} & 53.8 {\tiny[52.3, 55.3]} & 87.0 {\tiny[86.1, 87.9]} & 80.6 {\tiny[79.2, 82.1]} & 0.66 & 0.59 \\
Qwen 2.5 7B (local) & 81.1 {\tiny[79.9, 82.2]} & 52.6 {\tiny[51.1, 54.0]} & 83.8 {\tiny[82.7, 84.8]} & 77.3 {\tiny[75.7, 78.9]} & 0.61 & 0.58 \\
Claude Haiku 4.5 & 93.9 {\tiny[92.6, 95.0]} & 67.2 {\tiny[64.5, 69.8]} & 94.4 {\tiny[93.7, 95.1]} & 88.9 {\tiny[87.6, 90.2]} & 0.79 & 0.29 \\
\bottomrule
\end{tabular}
\end{table}

\begin{table}[!htbp]
\centering
\caption{Per-group coverage and total accuracy (percent), Study 2 four-category sample ($N = 17{,}998$). Coverage is the share of the group's voters the tool labels. Total accuracy is the share of ALL voters in the group who receive the correct label, so unlabeled voters count as errors. Black precision is the share of voters labeled Black who report Black identity, and Black recall the share of Black voters labeled Black.}
\label{tab:study2_total}
\footnotesize
\begin{tabular}{lcccccccccc}
\toprule
 & \multicolumn{4}{c}{Coverage} & \multicolumn{4}{c}{Total accuracy} & \multicolumn{2}{c}{Black} \\
\cmidrule(lr){2-5}\cmidrule(lr){6-9}\cmidrule(lr){10-11}
Tool & W & B & H & A & W & B & H & A & Prec. & Recall \\
\midrule
pyethnicity & 100.0 & 100.0 & 100.0 & 100.0 & 87.7 & 64.1 & 84.8 & 76.5 & 81.0 & 64.1 \\
WRU (+tract) & 99.9 & 99.6 & 99.7 & 99.2 & 88.0 & 62.5 & 78.1 & 62.1 & 77.4 & 62.5 \\
rethnicity & 100.0 & 100.0 & 100.0 & 100.0 & 66.3 & 71.5 & 79.2 & 80.5 & 66.1 & 71.5 \\
WRU (+county) & 99.8 & 99.2 & 99.8 & 99.3 & 85.3 & 54.3 & 78.2 & 59.4 & 70.4 & 54.3 \\
ethnicolr & 100.0 & 100.0 & 100.0 & 100.0 & 95.1 & 37.0 & 81.3 & 51.4 & 82.7 & 37.0 \\
WRU (surname) & 100.0 & 99.9 & 99.9 & 99.9 & 95.7 & 16.6 & 79.2 & 61.4 & 83.0 & 16.6 \\
predictrace & 93.8 & 94.8 & 69.8 & 81.0 & 90.1 & 15.2 & 53.8 & 59.4 & 84.6 & 15.2 \\
\midrule
Gemini 2.5 Flash & 89.5 & 82.6 & 97.8 & 94.0 & 81.4 & 44.4 & 85.1 & 75.8 & 85.8 & 44.4 \\
Qwen 2.5 7B (local) & 96.9 & 93.8 & 98.0 & 92.7 & 78.5 & 49.3 & 82.1 & 71.7 & 71.3 & 49.3 \\
Claude Haiku 4.5 & 32.5 & 25.0 & 81.8 & 70.6 & 30.5 & 16.8 & 77.2 & 62.8 & 95.8 & 16.8 \\
\bottomrule
\end{tabular}
\end{table}

\newpage
\section{Paired Comparisons and Matched Coverage in Study 2}
\label{app:paired}

Conditional accuracy scores each tool on the voters it chooses to label. Table~\ref{tab:paired} scores pairs of tools on the voters both label. Haiku labels a subset of the voters every full-coverage tool labels, so these comparisons use Haiku's labeled set. On it, pyethnicity is correct on 90.0\% and Haiku on 89.3\%, and Gemini is 0.4 points ahead of Haiku. Haiku is ahead of every Census-based tool, by 6.5 points over WRU with tract priors and 11.3 points over surname-only WRU.

Table~\ref{tab:matched} asks how accurate a probabilistic tool would be if it abstained as often as an LLM. Each tool keeps its most confident voters, ranked by the probability of its most likely category, until it labels as many voters as the LLM. At Haiku's coverage of 50.5\%, pyethnicity is correct on 94.1\%, 4.8 points more than Haiku, and WRU with tract priors on 88.4\%. At Gemini's and Qwen's higher coverage, pyethnicity is again more accurate than the LLM. Abstention by confidence raises the accuracy of every tool, and the LLMs' high conditional accuracy is largely the accuracy of abstention. Matching coverage does not match the composition of the labeled voters. Pyethnicity's most confident half contains 29.4\% of Black voters and Haiku's labeled half 25.0\%.

\begin{table}[!htbp]
\centering
\caption{Paired comparison on voters both tools label, Study 2 four-category sample. Accuracy is the share correct among the voters in the row, so both tools are scored on the same cases. The difference is the LLM minus the other tool, in percentage points, with 95\% stratified bootstrap intervals.}
\label{tab:paired}
\footnotesize
\begin{tabular}{lcccc}
\toprule
Compared with & $N$ both label & LLM accuracy & Tool accuracy & Difference (pp) \\
\midrule
\multicolumn{5}{l}{\textit{Claude Haiku 4.5}} \\
\quad WRU (surname) & 9,087 & 89.3 & 77.9 & $+11.3$ [$+10.6$, $+12.0$] \\
\quad WRU (+county) & 9,060 & 89.3 & 81.0 & $+8.3$ [$+7.5$, $+9.0$] \\
\quad WRU (+tract) & 9,059 & 89.3 & 82.9 & $+6.5$ [$+5.7$, $+7.2$] \\
\quad predictrace & 7,177 & 87.8 & 79.1 & $+8.7$ [$+8.0$, $+9.4$] \\
\quad rethnicity & 9,088 & 89.3 & 85.5 & $+3.8$ [$+3.2$, $+4.4$] \\
\quad ethnicolr & 9,088 & 89.3 & 82.3 & $+7.0$ [$+6.5$, $+7.6$] \\
\quad pyethnicity & 9,088 & 89.3 & 90.0 & $-0.8$ [$-1.1$, $-0.4$] \\
\quad Gemini 2.5 Flash & 9,024 & 89.5 & 89.9 & $-0.4$ [$-0.6$, $-0.1$] \\
\quad Qwen 2.5 7B & 8,941 & 89.3 & 87.9 & $+1.4$ [$+1.0$, $+1.8$] \\
\addlinespace
\multicolumn{5}{l}{\textit{Gemini 2.5 Flash}} \\
\quad WRU (surname) & 16,311 & 78.6 & 66.1 & $+12.5$ [$+11.9$, $+13.1$] \\
\quad WRU (+county) & 16,250 & 78.7 & 72.2 & $+6.5$ [$+5.8$, $+7.1$] \\
\quad WRU (+tract) & 16,266 & 78.6 & 75.3 & $+3.3$ [$+2.7$, $+3.9$] \\
\quad predictrace & 13,799 & 76.5 & 66.2 & $+10.3$ [$+9.7$, $+10.9$] \\
\quad rethnicity & 16,317 & 78.6 & 75.4 & $+3.2$ [$+2.6$, $+3.8$] \\
\quad ethnicolr & 16,317 & 78.6 & 70.4 & $+8.1$ [$+7.6$, $+8.7$] \\
\quad pyethnicity & 16,317 & 78.6 & 80.5 & $-1.9$ [$-2.4$, $-1.5$] \\
\quad Claude Haiku 4.5 &  9,024 & 89.9 & 89.5 & $+0.4$ [$+0.1$, $+0.6$] \\
\quad Qwen 2.5 7B & 15,773 & 78.8 & 75.8 & $+2.9$ [$+2.4$, $+3.4$] \\
\addlinespace
\multicolumn{5}{l}{\textit{Qwen 2.5 7B}} \\
\quad WRU (surname) & 17,203 & 73.5 & 64.5 & $+9.0$ [$+8.4$, $+9.6$] \\
\quad WRU (+county) & 17,136 & 73.5 & 71.5 & $+2.0$ [$+1.3$, $+2.7$] \\
\quad WRU (+tract) & 17,152 & 73.5 & 74.9 & $-1.4$ [$-2.1$, $-0.7$] \\
\quad predictrace & 14,785 & 71.2 & 64.1 & $+7.1$ [$+6.4$, $+7.8$] \\
\quad rethnicity & 17,212 & 73.5 & 74.5 & $-1.1$ [$-1.7$, $-0.5$] \\
\quad ethnicolr & 17,212 & 73.5 & 68.9 & $+4.6$ [$+4.0$, $+5.2$] \\
\quad pyethnicity & 17,212 & 73.5 & 78.8 & $-5.4$ [$-5.9$, $-4.8$] \\
\quad Claude Haiku 4.5 &  8,941 & 87.9 & 89.3 & $-1.4$ [$-1.8$, $-1.0$] \\
\quad Gemini 2.5 Flash & 15,773 & 75.8 & 78.8 & $-2.9$ [$-3.4$, $-2.4$] \\
\addlinespace
\bottomrule
\end{tabular}
\end{table}

\begin{table}[!htbp]
\centering
\caption{Accuracy at matched coverage, Study 2 four-category sample ($N = 17{,}998$). Each probabilistic tool labels only its most confident voters, ranked by the probability of its most likely category, in the same number as the LLM labels. The difference is the LLM's conditional accuracy minus the tool's, in percentage points, with 95\% stratified bootstrap intervals in which the most confident share is recomputed in every replicate.}
\label{tab:matched}
\small
\begin{tabular}{lcccc}
\toprule
 & Accuracy, & Accuracy at & LLM minus & Coverage of \\
Tool & all voters & matched coverage & tool (pp) & Black voters \\
\midrule
\multicolumn{5}{l}{\textit{Claude Haiku 4.5: coverage 50.5\%, conditional accuracy 89.3\%}} \\
\quad WRU (surname) & 63.5 & 85.5 & $+3.8$ [$+3.0$, $+4.6$] & 14.7 \\
\quad WRU (+county) & 70.7 & 87.0 & $+2.2$ [$+1.5$, $+3.0$] & 17.3 \\
\quad WRU (+tract) & 74.1 & 88.4 & $+0.9$ [$+0.1$, $+1.6$] & 25.1 \\
\quad pyethnicity & 78.5 & 94.1 & $-4.8$ [$-5.4$, $-4.2$] & 29.4 \\
\addlinespace
\multicolumn{5}{l}{\textit{Gemini 2.5 Flash: coverage 90.7\%, conditional accuracy 78.6\%}} \\
\quad WRU (surname) & 63.5 & 66.5 & $+12.0$ [$+11.5$, $+12.6$] & 89.6 \\
\quad WRU (+county) & 70.7 & 74.6 & $+3.9$ [$+3.2$, $+4.6$] & 88.4 \\
\quad WRU (+tract) & 74.1 & 77.7 & $+0.9$ [$+0.2$, $+1.6$] & 86.9 \\
\quad pyethnicity & 78.5 & 81.8 & $-3.2$ [$-3.7$, $-2.7$] & 83.5 \\
\addlinespace
\multicolumn{5}{l}{\textit{Qwen 2.5 7B: coverage 95.6\%, conditional accuracy 73.5\%}} \\
\quad WRU (surname) & 63.5 & 65.2 & $+8.3$ [$+7.6$, $+8.9$] & 98.0 \\
\quad WRU (+county) & 70.7 & 72.8 & $+0.6$ [$-0.1$, $+1.3$] & 97.2 \\
\quad WRU (+tract) & 74.1 & 75.9 & $-2.5$ [$-3.2$, $-1.8$] & 96.1 \\
\quad pyethnicity & 78.5 & 80.1 & $-6.7$ [$-7.2$, $-6.1$] & 93.0 \\
\addlinespace
\bottomrule
\end{tabular}
\end{table}

\newpage
\section{Population-Weighted Accuracy in Study 2}
\label{app:popweight}

Study~2 oversamples minority voters. Reweighting per-group accuracy to U.S.\ Census proportions favors tools that do well on White names (Table~\ref{tab:popweight}). Surname-only tools gain about 17 percentage points and geography-augmented WRU about 7 to 8. Rethnicity is the only tool that loses, because it trades accuracy on White names for accuracy on minority names.

\begin{table}[!htbp]
\centering
\caption{Four-category conditional accuracy under the stratified sample's own composition and under U.S.\ Census population weights (61\% White, 13\% Black, 20\% Hispanic, 6\% Asian), Study 2 ($N = 17{,}998$ names with a four-category self-report).}
\label{tab:popweight}
\small
\begin{tabular}{lccc}
\toprule
Tool & Sample-weighted & Population-weighted & Difference (pp) \\
\midrule
pyethnicity & 78.5 & 83.4 & +4.9 \\
WRU (+tract) & 74.1 & 81.3 & +7.2 \\
rethnicity & 73.7 & 70.4 & -3.3 \\
WRU (+county) & 70.7 & 78.5 & +7.8 \\
ethnicolr & 67.9 & 82.2 & +14.3 \\
WRU (surname) & 63.5 & 80.1 & +16.6 \\
predictrace & 63.5 & 80.5 & +17.1 \\
\midrule
Claude Haiku 4.5 & 89.3 & 90.2 & +0.9 \\
Gemini 2.5 Flash & 78.6 & 84.7 & +6.1 \\
Qwen 2.5 7B (local) & 73.5 & 77.7 & +4.2 \\
\midrule
Hybrid Haiku$\rightarrow$pyethnicity & 78.1 & 83.5 & +5.5 \\
\bottomrule
\end{tabular}
\end{table}

\newpage
\section{Calibration of WRU}
\label{app:calibration}

Where WRU's most likely category has a probability above 80\%, it is correct about as often as the probability says (Table~\ref{tab:calibration_wru}). Between 40\% and 80\% it is overconfident. For the 4,102 voters in the 40 to 60\% bin, its most likely category is correct for 35.9\% at a mean predicted probability of 49.4\%.

\begin{table}[!htbp]
\centering
\caption{Calibration of WRU (surname) by the probability of its most likely category, Study 2 four-category sample ($N = 17{,}998$). Accuracy is the share of names whose most likely category matches the self-report.}
\label{tab:calibration_wru}
\small
\begin{tabular}{lccc}
\toprule
Probability bin & $N$ & Accuracy & Mean predicted probability \\
\midrule
Below 40\% &    53 & 39.6\% & 36.8\% \\
40--60\% & 4,102 & 35.9\% & 49.4\% \\
60--80\% & 5,551 & 47.6\% & 69.3\% \\
80--90\% & 2,747 & 77.4\% & 85.8\% \\
90--100\% & 5,545 & 93.0\% & 93.9\% \\
\bottomrule
\end{tabular}
\end{table}

\newpage
\section{The Residual Category}
\label{app:other}

Of the 2,000 sampled voters outside the four categories, 60\% report ``Other'' race, 26\% American Indian, and 14\% multiracial. No tool can return these categories, so every label they receive is wrong under the four-category scheme. The Census tools assign most of them to White (71\% of the labels surname-only WRU gives them and 69\% of predictrace's, against 50\% with county priors). The LLMs leave many unlabeled (Haiku 65\%). The four-category scheme has no place for these voters, and an analysis that uses name-based classification should state how it treats them.

\newpage
\section{Refusal Analysis and Hybrid Pipelines}
\label{app:refusal}
\label{app:hybrid}

A hybrid that fills Haiku's missing labels with pyethnicity would beat both components if Haiku's refusals were unrelated to name difficulty. They are not. Pyethnicity is correct on 90.0\% of the voters Haiku labels and on 66.6\% of those it leaves unlabeled, a gap of 23.4 points (95\% interval 22.3 to 24.5), and the gap appears in every group (Table~\ref{tab:refusal}). Figure~\ref{fig:pipeline} shows the consequence across population compositions. Under the assumption of random refusal the hybrid would reach 86.8\% total accuracy at U.S.\ Census proportions. With the actual refusals it reaches 83.5\%, against 83.4\% for pyethnicity alone.

\begin{table}[!htbp]
\centering
\caption{Accuracy of pyethnicity on the voters Claude Haiku labels and on those it leaves unlabeled, Study 2 four-category sample. If Haiku's refusals were unrelated to name difficulty, the two rows would be similar.}
\label{tab:refusal}
\small
\begin{tabular}{lcccccc}
\toprule
Voters & $N$ & Overall & White & Black & Hispanic & Asian \\
\midrule
Labeled by Haiku &  9,088 & 90.0 & 91.8 & 76.5 & 94.1 & 88.9 \\
Left unlabeled by Haiku &  8,910 & 66.6 & 85.7 & 59.9 & 43.1 & 46.6 \\
All voters & 17,998 & 78.5 & 87.7 & 64.1 & 84.8 & 76.5 \\
\bottomrule
\end{tabular}
\end{table}

\begin{figure}[htbp]
\centering
\includegraphics[width=\textwidth]{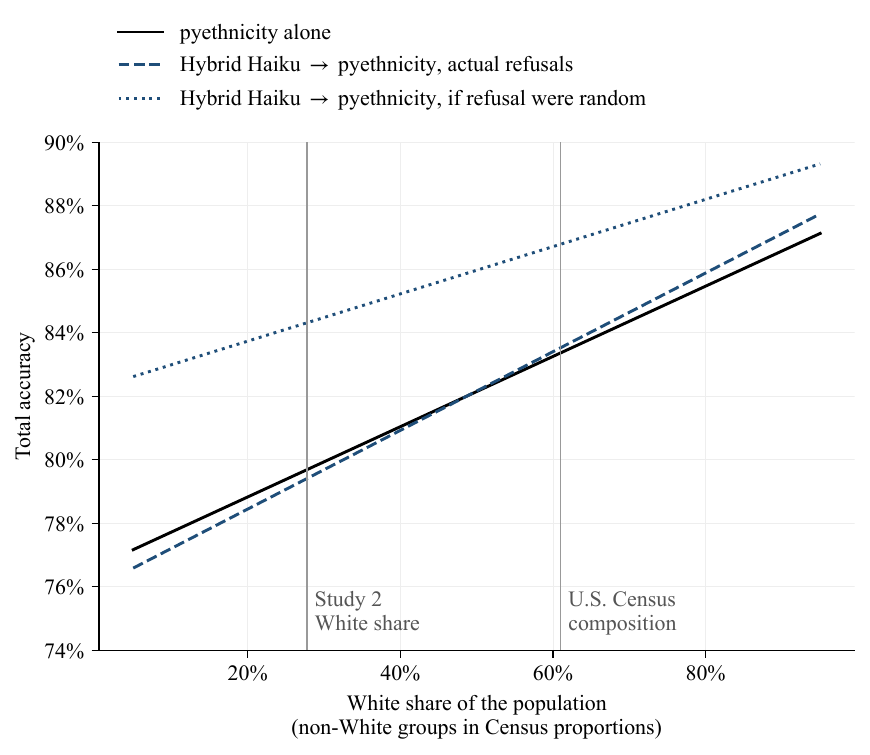}
\caption{Total accuracy of pyethnicity alone and of the Haiku and pyethnicity hybrid as the White share of the population varies, with non-White groups in Census proportions. The dotted line shows what the hybrid would achieve if Haiku's refusals were unrelated to name difficulty. Based on the 17,998 four-category Study 2 voters.}
\label{fig:pipeline}
\end{figure}

\newpage
\section{Study 3 Design and Additional Results}
\label{app:study3full}
\label{app:strata}

The Study~3 sample was stratified to oversample informative cases, using WRU's predictions and an earlier LLM run. S1 and S2 contain authors on whom WRU and the LLM agreed (White and non-White). S3 and S4 contain authors on whom they disagreed. S5 contains authors for whom no WRU category exceeds 50\%. S6 contains authors for whom the earlier LLM run reported low confidence. S7 contains authors labeled only by traditional tools. S8 contains authors for whom only the LLM predicted gender. S10 is a non-White oversample, and no authors were drawn under the label S9. The strata contain 40, 36, 30, 20, 50, 20, 30, 40, and 34 authors.

Because the strata are defined by the tools' own outputs, unweighted accuracy describes performance on hard cases. Table~\ref{tab:study3_weighted} also reports accuracy weighted to the sampling frame. Weighted, surname-only WRU and Haiku are about equally accurate, and the LLMs' advantage in the unweighted comparison comes mostly from the oversampled hard cases. Scored on the 101 authors both label, Haiku is correct on 90.1\% and WRU on 80.2\% unweighted, and on 97.5\% and 94.5\% weighted. Among authors whom both models labeled, Haiku and Gemini agree on 93.9\% of race labels and 98.9\% of gender labels.

The reference codes come from one coder who inferred race and gender from public profiles. A second coder independently coded 75 authors for the reliability figures in Section~\ref{sec:study3}. Only the agreement statistics of that exercise are available to us, not the second coder's codes for each author, so we cannot rescore the tools against the second coder. As a sensitivity analysis of category boundaries, the lenient column of Table~\ref{tab:study3_weighted} counts a label as correct when it and the reference code differ only across the Asian and White or the Hispanic and White boundary. Lenient scoring raises every tool, most of all the LLMs (Haiku from 89.5\% to 97.4\%), because most LLM errors in this sample lie at these boundaries. Of the 13 authors coded as other or mixed race, WRU labels 8, Gemini 3, and Haiku 2, and every such label is wrong under the four-category scheme.

\begin{table}[!htbp]
\centering
\caption{Study 3 race accuracy against human coding, unweighted and weighted to the sampling frame. The sample oversamples strata defined by agreement between WRU and an earlier LLM run, so unweighted accuracy describes performance on hard cases. Design weights are stratum frame counts divided by stratum sample sizes, and the non-White oversample receives weight zero. Lenient scoring also counts a label as correct when it and the reference code differ only across the Asian and White or the Hispanic and White boundary. This is a sensitivity analysis, not rescoring against the second coder.}
\label{tab:study3_weighted}
\small
\begin{tabular}{lccccccc}
\toprule
 & \multicolumn{4}{c}{Four categories} & \multicolumn{3}{c}{White vs.\ non-White} \\
\cmidrule(lr){2-5}\cmidrule(lr){6-8}
Tool & $N$ & Unweighted & Weighted & Lenient & $N$ & Unweighted & Weighted \\
\midrule
ethnicolr & 224 & 74.6 & 82.9 & 85.7 & 237 & 77.6 & 85.6 \\
pyethnicity & 224 & 71.0 & 78.3 & 83.9 & 237 & 71.3 & 79.5 \\
predictrace & 160 & 79.4 & 92.6 & 86.2 & 169 & 77.5 & 92.1 \\
rethnicity & 224 & 55.8 & 61.6 & 79.5 & 237 & 65.0 & 67.9 \\
WRU (surname) & 152 & 80.3 & 94.6 & 86.8 & 160 & 78.1 & 93.9 \\
\midrule
Gemini 2.5 Flash & 203 & 88.7 & 91.7 & 97.0 & 228 & 86.8 & 90.2 \\
Qwen 2.5 7B (local) & 173 & 83.8 & 91.4 & 93.1 & 207 & 82.1 & 89.5 \\
Claude Haiku 4.5 & 153 & 89.5 & 94.3 & 97.4 & 175 & 89.1 & 93.1 \\
\bottomrule
\end{tabular}
\end{table}

\begin{figure}[htbp]
\centering
\includegraphics[width=\textwidth]{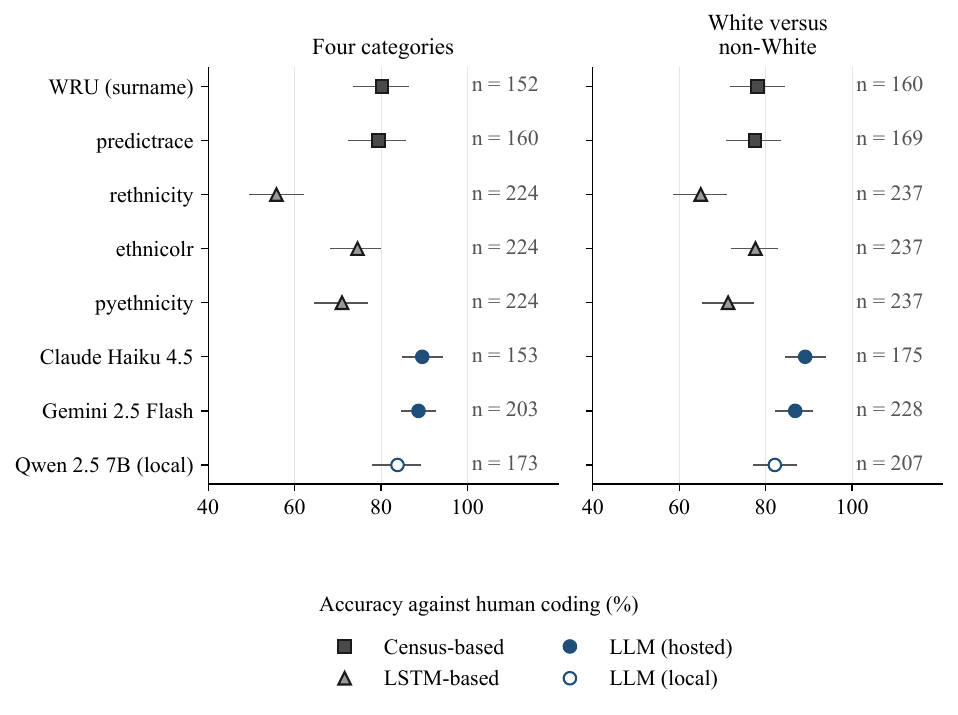}
\caption{Unweighted Study 3 accuracy against human coding, four categories and White versus non-White. Lines are 95\% bootstrap intervals, and $n$ is the number of authors whom both the tool and the coder classified.}
\label{fig:study3_comparison}
\end{figure}

\newpage
\section{Throughput and Cost}
\label{app:scalability}

Traditional tools label tens of thousands of names in seconds or minutes. The hosted LLMs classified about 15 (Haiku) and 30 (Gemini) names per second with parallel requests, so 100,000 names take about one to two hours. Batch interfaces reduce this further. API fees for the hosted models in this study came to about \$28 for roughly 40,000 names per model, or about \$0.45 (Haiku) and \$0.25 (Gemini) per thousand names. The local model classified about 4 to 5 names per second on a laptop and costs nothing per name. These figures exclude researcher time for prompt design, parsing, and validation.

\newpage
\section{Empirical Confusion Matrices and Differential-Error Diagnostics}
\label{app:confusion}

Table~\ref{app:tab_confusion} reports the confusion matrices of the tools featured in Section~\ref{sec:error_propagation}, tabulated from the predicted and self-reported labels of all 19,998 Study~2 voters. Each row sums to one including the final column, which records voters who received no label. Figure~\ref{app:fig_confusion} displays the same matrices.

\begin{table}[!htbp]
\centering
\caption{Empirical confusion matrices, Study 2. Each cell is the row-conditional probability $\Pr(\hat{R}=j \mid R=i)$ estimated by the sample share. Rows sum to one including the final column, which records names the tool left unclassified.}
\label{app:tab_confusion}
\footnotesize
\begin{tabular}{lccccc}
\toprule
True race & White & Black & Hispanic & Asian & Unclassified \\
\midrule
\multicolumn{6}{l}{\textit{WRU (surname)}} \\
White ($N{=}$5,000) & 0.957 & 0.023 & 0.014 & 0.005 & 0.000 \\
Black ($N{=}$5,000) & 0.820 & 0.166 & 0.006 & 0.007 & 0.001 \\
Hispanic ($N{=}$5,000) & 0.192 & 0.006 & 0.792 & 0.010 & 0.001 \\
Asian ($N{=}$2,998) & 0.342 & 0.009 & 0.033 & 0.614 & 0.001 \\
Other ($N{=}$2,000) & 0.671 & 0.050 & 0.100 & 0.127 & 0.052 \\
\addlinespace
\multicolumn{6}{l}{\textit{WRU (+county)}} \\
White ($N{=}$5,000) & 0.853 & 0.128 & 0.013 & 0.004 & 0.002 \\
Black ($N{=}$5,000) & 0.440 & 0.543 & 0.006 & 0.004 & 0.008 \\
Hispanic ($N{=}$5,000) & 0.165 & 0.042 & 0.782 & 0.009 & 0.002 \\
Asian ($N{=}$2,998) & 0.269 & 0.099 & 0.031 & 0.594 & 0.007 \\
Other ($N{=}$2,000) & 0.441 & 0.228 & 0.097 & 0.118 & 0.117 \\
\addlinespace
\multicolumn{6}{l}{\textit{pyethnicity}} \\
White ($N{=}$5,000) & 0.877 & 0.093 & 0.022 & 0.008 & 0.000 \\
Black ($N{=}$5,000) & 0.340 & 0.641 & 0.012 & 0.007 & 0.000 \\
Hispanic ($N{=}$5,000) & 0.107 & 0.035 & 0.848 & 0.009 & 0.000 \\
Asian ($N{=}$2,998) & 0.134 & 0.036 & 0.065 & 0.765 & 0.000 \\
Other ($N{=}$2,000) & 0.455 & 0.242 & 0.124 & 0.179 & 0.000 \\
\addlinespace
\multicolumn{6}{l}{\textit{Claude Haiku 4.5}} \\
White ($N{=}$5,000) & 0.305 & 0.003 & 0.013 & 0.003 & 0.675 \\
Black ($N{=}$5,000) & 0.069 & 0.168 & 0.010 & 0.003 & 0.750 \\
Hispanic ($N{=}$5,000) & 0.041 & 0.003 & 0.772 & 0.002 & 0.182 \\
Asian ($N{=}$2,998) & 0.036 & 0.001 & 0.041 & 0.628 & 0.294 \\
Other ($N{=}$2,000) & 0.097 & 0.037 & 0.098 & 0.121 & 0.647 \\
\addlinespace
\multicolumn{6}{l}{\textit{Gemini 2.5 Flash}} \\
White ($N{=}$5,000) & 0.814 & 0.051 & 0.024 & 0.006 & 0.105 \\
Black ($N{=}$5,000) & 0.348 & 0.444 & 0.024 & 0.010 & 0.174 \\
Hispanic ($N{=}$5,000) & 0.107 & 0.015 & 0.851 & 0.005 & 0.022 \\
Asian ($N{=}$2,998) & 0.110 & 0.014 & 0.058 & 0.758 & 0.060 \\
Other ($N{=}$2,000) & 0.365 & 0.143 & 0.124 & 0.164 & 0.203 \\
\addlinespace
\bottomrule
\end{tabular}
\end{table}

\begin{figure}[htbp]
\centering
\includegraphics[width=\textwidth]{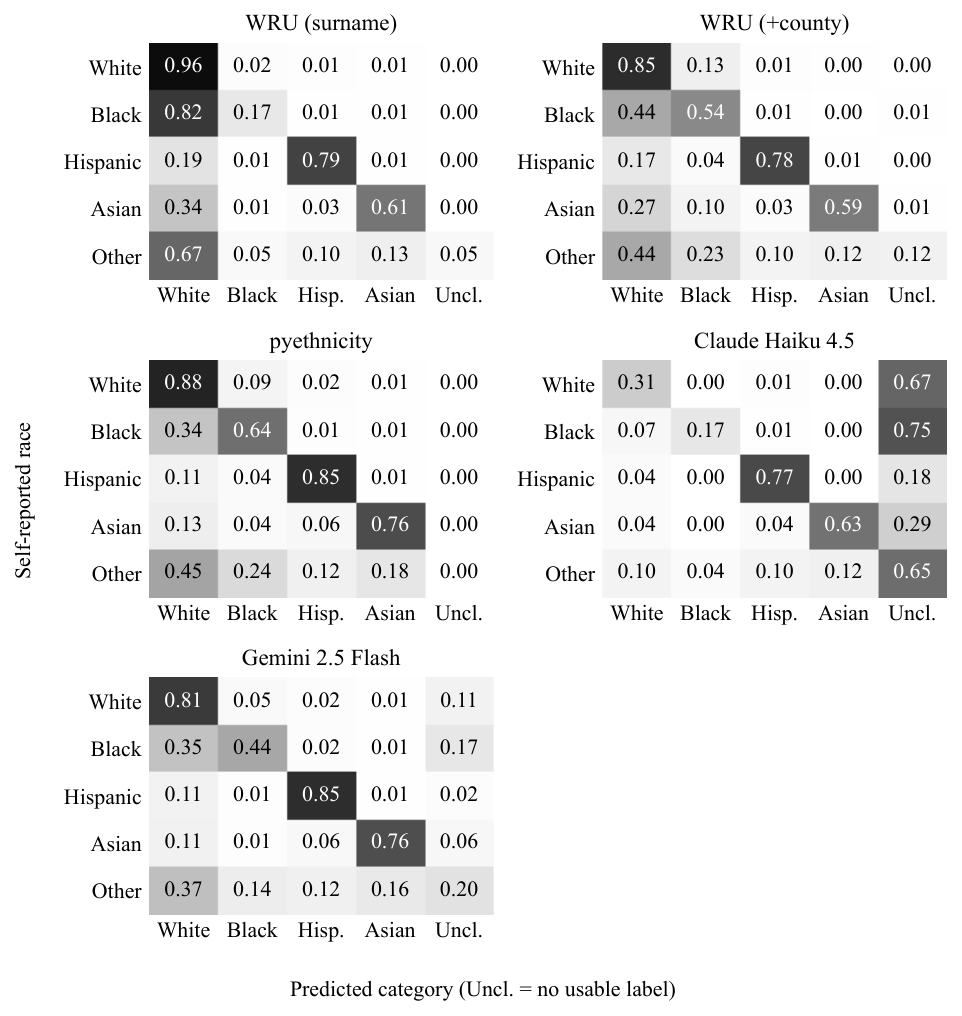}
\caption{Empirical confusion matrices, Study 2 ($N = 19{,}998$). Each cell is the share of voters in the row's self-reported group who receive the column's label.}
\label{app:fig_confusion}
\end{figure}

Table~\ref{tab:nondifferential} compares, for voters each tool labels White or Black, their observed mean age with the mean age expected if labeling were unrelated to age given self-reported race. For the traditional tools the gaps are within about 1.7 years. Voters Haiku labels Black are 7.4 years younger than expected, which is the selection that drives Haiku's bias in the age gap. Gemini ($-3.5$ years) and Qwen ($-2.6$ years) show smaller gaps in the same direction, but because they label nearly every voter, their gaps come from misclassification rather than from refusal.

\begin{table}[!htbp]
\centering
\caption{Differential-error diagnostic for age, Study 2. For voters a tool labels White or Black, the table compares their mean age with the mean age expected if labeling were unrelated to age given self-reported race (the group means weighted by the true composition of the labeled set). A gap far from zero means that the tool's labeling selects on the outcome within racial groups.}
\label{tab:nondifferential}
\small
\begin{tabular}{lcccccc}
\toprule
 & \multicolumn{3}{c}{Labeled White} & \multicolumn{3}{c}{Labeled Black} \\
\cmidrule(lr){2-4}\cmidrule(lr){5-7}
Tool & Expected & Observed & Gap & Expected & Observed & Gap \\
\midrule
WRU (surname) & 49.8 & 50.1 & +0.3 & 49.7 & 48.8 & -0.8 \\
WRU (+county) & 50.0 & 50.4 & +0.4 & 49.2 & 48.7 & -0.5 \\
WRU (+tract) & 50.1 & 50.4 & +0.4 & 49.3 & 48.8 & -0.5 \\
pyethnicity & 50.8 & 51.6 & +0.8 & 49.4 & 47.8 & -1.6 \\
rethnicity & 50.9 & 52.0 & +1.1 & 49.8 & 48.8 & -1.0 \\
Claude Haiku 4.5 & 51.2 & 52.6 & +1.4 & 49.6 & 42.3 & -7.4 \\
Gemini 2.5 Flash & 50.8 & 51.8 & +1.0 & 49.6 & 46.1 & -3.5 \\
Qwen 2.5 7B (local) & 50.7 & 51.8 & +1.2 & 49.7 & 47.0 & -2.6 \\
Hybrid Haiku$\rightarrow$pyethnicity & 50.8 & 51.6 & +0.9 & 49.4 & 47.5 & -2.0 \\
\bottomrule
\end{tabular}
\end{table}

Table~\ref{tab:missing_type} separates the selection term by the type of missing output. For Haiku, almost all of the selection bias in the age gap comes from its ``Unknown'' answers. Dropping only those voters inflates the age gap by 3.4 years, because the Black voters Haiku answers ``Unknown'' for are on average 55.0 years old, against 50.1 for all Black voters. Dropping only its explicit refusals moves the gap by $-1.3$ years in the other direction, since the Black voters it refuses are younger (47.5 years). The answers outside the four categories are too few to matter. The output types locate the selection on age but do not identify whether uncertainty or safety training explains it. For the registration gap the selection terms are small for every model and every type.

\begin{table}[!htbp]
\centering
\caption{Selection bias by type of missing LLM output, Study 2. Each row drops only the voters with that type of missing output and computes the gap from self-reported race among the remaining voters. The difference from the true gap is the selection bias that type of missing output causes on its own. The last row of each panel drops all unlabeled voters and equals the selection term of Equation~\ref{eq:decomposition}. The parts need not add up to it. Black means are among self-reported Black voters with that type of missing output. Brackets give 95\% stratified bootstrap intervals.}
\label{tab:missing_type}
\scriptsize
\begin{tabular}{lccccc}
\toprule
 & & \multicolumn{2}{c}{Democratic registration} & \multicolumn{2}{c}{Age} \\
\cmidrule(lr){3-4}\cmidrule(lr){5-6}
Dropped voters & $N$ (age) & Black \% Dem & Selection bias & Black mean age & Selection bias \\
\midrule
\multicolumn{6}{l}{\textit{Claude Haiku 4.5}} \\
\quad Explicit refusal & 4,040 & 71.3 & $-0.008$ [$-0.018$, $+0.002$] & 47.5 & $-1.28$ [$-1.73$, $-0.80$] \\
\quad ``Unknown'' & 5,405 & 76.4 & $+0.035$ [$+0.019$, $+0.050$] & 55.0 & $+3.36$ [$+2.63$, $+4.09$] \\
\quad Outside categories &   224 & 47.1 & $-0.002$ [$-0.003$, $-0.000$] & 43.9 & $+0.00$ [$-0.05$, $+0.05$] \\
\quad All unlabeled & 9,669 & 74.0 & $+0.042$ [$+0.015$, $+0.067$] & 51.5 & $+4.39$ [$+3.26$, $+5.66$] \\
\addlinespace
\multicolumn{6}{l}{\textit{Gemini 2.5 Flash}} \\
\quad Explicit refusal &     1 & NA & $-0.000$ [$-0.001$, $+0.000$] & NA & $+0.00$ [$+0.00$, $+0.01$] \\
\quad ``Unknown'' & 1,618 & 77.5 & $+0.009$ [$+0.002$, $+0.015$] & 52.4 & $+0.22$ [$-0.08$, $+0.52$] \\
\quad Outside categories &   301 & 56.8 & $-0.002$ [$-0.004$, $-0.000$] & 45.2 & $+0.02$ [$-0.04$, $+0.09$] \\
\quad All unlabeled & 1,920 & 76.5 & $+0.006$ [$-0.001$, $+0.012$] & 52.0 & $+0.25$ [$-0.05$, $+0.56$] \\
\addlinespace
\multicolumn{6}{l}{\textit{Qwen 2.5 7B}} \\
\quad Explicit refusal &    46 & 56.0 & $-0.001$ [$-0.002$, $+0.000$] & 47.1 & $-0.01$ [$-0.05$, $+0.04$] \\
\quad ``Unknown'' &   485 & 73.8 & $+0.000$ [$-0.003$, $+0.003$] & 52.8 & $+0.06$ [$-0.08$, $+0.21$] \\
\quad Outside categories &   498 & 57.1 & $-0.003$ [$-0.006$, $-0.001$] & 40.7 & $-0.12$ [$-0.22$, $-0.02$] \\
\quad All unlabeled & 1,029 & 67.8 & $-0.004$ [$-0.009$, $-0.000$] & 48.9 & $-0.07$ [$-0.24$, $+0.12$] \\
\addlinespace
\bottomrule
\end{tabular}
\end{table}

\newpage
\section{Sensitivity of the Downstream Analysis}
\label{app:sensitivity}

\paragraph{Voters outside the four categories.} The main downstream analysis keeps the 2,000 voters who report Other, American Indian, or multiracial identity, because a researcher cannot remove them before classifying and tools can label them White or Black. The two gaps compare self-reported White and Black voters, so dropping these voters leaves the true gaps unchanged and changes only the estimators. Table~\ref{tab:downstream_fourcat} shows that the conclusions hold in the four-category sample. Hard labels understate the registration gap by 33\% to 61\% and posterior weighting by 55\% to 79\%, and Haiku's hard labels overstate the age gap by 6.5 years. The pooled BIRDiE estimates move from within 0.03 of the true gap to within 0.05 of it, because the posteriors still assign probability to a category with no members.

\begin{table}[!htbp]
\centering
\caption{Downstream bias with and without the 2,000 voters outside the four categories (Other, American Indian, and multiracial). The main analysis keeps them, so tools can mislabel them as White or Black. In the four-category sample they are dropped from both the estimators and the truth.}
\label{tab:downstream_fourcat}
\footnotesize
\begin{tabular}{lcccc}
\toprule
 & \multicolumn{2}{c}{Bias($\hat{\Delta}_{\text{party}}$)} & \multicolumn{2}{c}{Bias($\hat{\Delta}_{\text{age}}$, years)} \\
\cmidrule(lr){2-3}\cmidrule(lr){4-5}
Estimator & All voters & Four categories & All voters & Four categories \\
\midrule
\multicolumn{5}{l}{\textit{Hard-label (argmax) estimators}} \\
\quad WRU surname & $+0.334$ & $+0.320$ & $-3.07$ & $-3.40$ \\
\quad WRU +county & $+0.298$ & $+0.278$ & $-2.62$ & $-2.56$ \\
\quad WRU +tract & $+0.240$ & $+0.217$ & $-2.69$ & $-2.69$ \\
\quad pyethnicity & $+0.257$ & $+0.236$ & $-0.52$ & $-0.44$ \\
\quad Claude Haiku 4.5 & $+0.191$ & $+0.173$ & $+6.04$ & $+6.46$ \\
\quad Gemini 2.5 Flash & $+0.271$ & $+0.249$ & $+1.42$ & $+1.64$ \\
\quad Qwen 2.5 7B (local) & $+0.344$ & $+0.325$ & $+0.53$ & $+0.59$ \\
\quad Hybrid Haiku$\rightarrow$pyeth. & $+0.262$ & $+0.241$ & $-0.11$ & $+0.01$ \\
\addlinespace
\multicolumn{5}{l}{\textit{Posterior-weighting estimators}} \\
\quad WRU surname & $+0.426$ & $+0.416$ & $-3.82$ & $-3.90$ \\
\quad WRU +county & $+0.367$ & $+0.353$ & $-3.43$ & $-3.41$ \\
\quad BISG tract chain & $+0.307$ & $+0.290$ & $-3.32$ & $-3.33$ \\
\quad pyethnicity & $+0.323$ & $+0.305$ & $-1.03$ & $-0.92$ \\
\addlinespace
\multicolumn{5}{l}{\textit{Mixed LLM + posterior pipelines}} \\
\quad Haiku + pyeth.\ posterior & $+0.316$ & $+0.297$ & $-0.47$ & $-0.33$ \\
\quad Haiku + BISG posterior & $+0.283$ & $+0.265$ & $-1.38$ & $-1.23$ \\
\addlinespace
\multicolumn{5}{l}{\textit{BIRDiE estimators}} \\
\quad BIRDiE, WRU surname & $-0.029$ & $-0.044$ & $+1.23$ & $+1.25$ \\
\quad BIRDiE, WRU +county & $-0.012$ & $-0.043$ & $+0.60$ & $+1.22$ \\
\quad BIRDiE, BISG tract chain & $-0.020$ & $-0.049$ & $-0.53$ & $-0.25$ \\
\addlinespace
\bottomrule
\end{tabular}
\end{table}

\paragraph{BIRDiE specification.} All BIRDiE models use the \texttt{birdie} R package (version 0.7.1) with the EM algorithm and the package's default prior. The categorical party model has four outcome levels (Democratic, Republican, unaffiliated, and the minor parties combined), and the race columns are the five BISG categories in the order White, Black, Hispanic, Asian, other. The main specification pools all counties (\texttt{party4 \textasciitilde{} 1}, \texttt{age \textasciitilde{} 1}). Table~\ref{tab:birdie_variants} adds county fixed effects (\texttt{\textasciitilde{} county}), county random effects for party (\texttt{\textasciitilde{} (1 | county)} with the mixed categorical family), and a categorical model for age deciles in place of the normal model. All BIRDiE fits reported here, including the bootstrap refits, met the package's convergence criterion. The registration gap is sensitive to the county specification when the posteriors use the surname alone or county priors, with biases up to $+0.120$, and stable when they use tract priors. The table also reports how informative each set of posteriors is and the range of the three party estimates, which shrinks as the posteriors sharpen. Restricting the tract models to the voters geocoded to a tract, so that no posterior falls back on county priors, leaves this pattern unchanged. For the age gap the specifications disagree, with biases between $-2.47$ and $+1.23$ years.

\begin{table}[!htbp]
\centering
\caption{BIRDiE bias under alternative specifications, Study 2. Pooled models assume one outcome distribution per race for the whole state (the main specification). County fixed effects let the distribution vary freely across the 100 counties, and county random effects shrink the county distributions toward a common mean. The sample has a median of 78.5 voters per county, and 37 counties have fewer than 50. The categorical age model replaces the normal model with ten age bins. $\bar{p}_{\text{true}}$ is the mean posterior probability of the self-reported race, and $\bar{p}_{B \mid B}$ the mean probability of Black among Black voters. Entries under each gap are biases, except Spread, which is the range of the three registration-gap estimates and requires no ground truth. The last row refits the tract models on the voters geocoded to a tract, against the true gap among them ($-0.517$).}
\label{tab:birdie_variants}
\scriptsize
\resizebox{\textwidth}{!}{%
\begin{tabular}{lccccccccc}
\toprule
 & \multicolumn{2}{c}{Posterior} & \multicolumn{4}{c}{Registration gap} & \multicolumn{3}{c}{Age gap (years)} \\
\cmidrule(lr){2-3}\cmidrule(lr){4-7}\cmidrule(lr){8-10}
Posterior & $\bar{p}_{\text{true}}$ & $\bar{p}_{B \mid B}$ & Pooled & County FE & County RE & Spread & Pooled & County FE & Age bins \\
\midrule
WRU surname & 0.54 & 0.31 & $-0.029$ & $+0.120$ & $+0.092$ & 0.149 & $+1.23$ & $-2.05$ & $-0.73$ \\
WRU +county & 0.57 & 0.48 & $-0.012$ & $+0.070$ & $+0.054$ & 0.082 & $+0.60$ & $-0.48$ & $-0.24$ \\
BISG tract chain & 0.60 & 0.55 & $-0.020$ & $+0.023$ & $+0.009$ & 0.043 & $-0.53$ & $-2.47$ & $-0.92$ \\
\quad geocoded voters only ($N = 15{,}799$) & --- & --- & $-0.028$ & $+0.014$ & $-0.001$ & 0.041 & --- & --- & --- \\
\bottomrule
\end{tabular}}
\end{table}

\paragraph{The independence assumption.} BIRDiE assumes that the outcome is independent of the name given race and geography. With self-reported race we can test for an association with a surname-based race probability. Table~\ref{tab:birdie_assumption} regresses each outcome on the surname-only probability of the Black category within self-reported White and Black voters, with county fixed effects. None of the four slopes is distinguishable from zero. The test has limited power, since surnames vary little in this probability within race (standard deviations of 0.13 among White and 0.18 among Black voters).

\begin{table}[!htbp]
\centering
\caption{Within-race association between outcomes and the surname-only Black probability, Study 2. Each row regresses the outcome on the probability that WRU assigns to the Black category from the surname alone, with county fixed effects, among voters of one self-reported race. BIRDiE assumes this slope is zero.}
\label{tab:birdie_assumption}
\small
\begin{tabular}{llcccc}
\toprule
Self-reported race & Outcome & $N$ & Slope & SE & $p$ \\
\midrule
White & Democratic registration & 4,764 & $-0.043$ & 0.045 & 0.34 \\
White & Age (years) & 4,690 & $+0.375$ & 2.317 & 0.87 \\
Black & Democratic registration & 4,876 & $-0.010$ & 0.036 & 0.79 \\
Black & Age (years) & 4,752 & $-2.041$ & 1.580 & 0.20 \\
\bottomrule
\end{tabular}
\end{table}

\paragraph{Comparing estimators.} Because every estimator is computed on the same bootstrap samples, differences in absolute bias between two estimators have paired intervals. Each BIRDiE estimate of the registration gap is closer to the truth than Haiku's hard labels in all 500 samples. Haiku's hybrid with a pyethnicity fallback is further from the true registration gap than Haiku alone in all 500 samples, by 0.072 on average (95\% interval 0.042 to 0.106), because the fallback adds pyethnicity's misclassification. For the age gap, the hybrid is closer to the truth than BIRDiE in 69\% to 88\% of the samples, depending on the posterior, and every interval for the difference contains zero.

\newpage
\section{Combining LLM Answers with BISG Posteriors}
\label{app:combination}

Let $p_{ij} = \Pr(R_i = j \mid S_i, G_i)$ be voter $i$'s BISG posterior and $L_i$ the LLM's answer, coded as one of the four categories or as the type of missing output (explicit refusal, ``Unknown'', or a category outside the four). The combined posterior is
\begin{equation*}
\tilde{p}_{ij} = \frac{p_{ij} \Pr(L_i \mid R = j)}{\sum_k p_{ik} \Pr(L_i \mid R = k)}.
\end{equation*}
We estimate $\Pr(L \mid R)$ as the share of voters with self-reported race $R$ who receive answer $L$, adding 0.5 to each count. The sample is split into two halves, stratified by self-reported race, and each half's posteriors use the rates estimated on the other half. In the bootstrap the halves are fixed per voter, so the copies of a voter drawn into one replicate share a half, and the rates are re-estimated in every replicate.

The update treats the LLM's answer as independent of the surname and geography given race. It is not, because the LLM reads the same surname, so the combined posteriors count the surname's evidence twice. They are sharper but overconfident. Table~\ref{tab:llm_bisg} shows that the combination raises the probability placed on the self-reported race and the accuracy of the most likely category for every posterior and every LLM, while the log loss rises in eight of the nine combinations, the signature of overconfidence. The exploratory check below instead estimates race probabilities conditional on the BISG posterior and the LLM's answer.

\begin{table}[!htbp]
\centering
\caption{LLM answers combined with BISG posteriors, Study 2. Each LLM's answer, including the type of missing output, updates the BISG posterior by Bayes' rule with $\Pr(L \mid R)$ estimated on the other half of the sample. Accuracy is the share of the four-category sample whose most likely category is the self-reported one (pyethnicity reaches 78.5\%). $\bar{p}_{\text{true}}$ is the mean probability of the self-reported race and $\bar{p}_{B \mid B}$ the mean probability of Black among Black voters. Biases are for the weighted estimator and pooled BIRDiE. True gaps are $-0.529$ (party) and 4.29 years (age).}
\label{tab:llm_bisg}
\scriptsize
\resizebox{\textwidth}{!}{%
\begin{tabular}{lcccccccc}
\toprule
 & \multicolumn{4}{c}{Classification} & \multicolumn{2}{c}{Weighted, bias} & \multicolumn{2}{c}{BIRDiE, bias} \\
\cmidrule(lr){2-5}\cmidrule(lr){6-7}\cmidrule(lr){8-9}
Posterior & Accuracy (\%) & $\bar{p}_{\text{true}}$ & Log loss & $\bar{p}_{B \mid B}$ & Party & Age & Party & Age \\
\midrule
\multicolumn{9}{l}{\textit{WRU surname}} \\
\quad alone & 63.5 & 0.536 & 1.131 & 0.309 & $+0.426$ & $-3.82$ & $-0.029$ & $+1.23$ \\
\quad + Haiku & 71.2 & 0.617 & 1.136 & 0.424 & $+0.374$ & $-0.90$ & $+0.099$ & $+5.83$ \\
\quad + Gemini & 75.5 & 0.640 & 1.127 & 0.466 & $+0.345$ & $-1.38$ & $+0.116$ & $+2.76$ \\
\quad + Qwen & 72.5 & 0.611 & 1.149 & 0.405 & $+0.381$ & $-2.05$ & $+0.082$ & $+3.29$ \\
\addlinespace
\multicolumn{9}{l}{\textit{WRU +county}} \\
\quad alone & 70.7 & 0.570 & 1.086 & 0.479 & $+0.367$ & $-3.43$ & $-0.012$ & $+0.60$ \\
\quad + Haiku & 76.2 & 0.648 & 1.101 & 0.562 & $+0.328$ & $-1.07$ & $+0.065$ & $+4.09$ \\
\quad + Gemini & 77.6 & 0.666 & 1.087 & 0.585 & $+0.305$ & $-1.32$ & $+0.083$ & $+2.16$ \\
\quad + Qwen & 75.8 & 0.644 & 1.104 & 0.548 & $+0.330$ & $-1.81$ & $+0.059$ & $+2.30$ \\
\addlinespace
\multicolumn{9}{l}{\textit{BISG tract chain}} \\
\quad alone & 74.1 & 0.600 & 1.047 & 0.546 & $+0.307$ & $-3.32$ & $-0.020$ & $-0.53$ \\
\quad + Haiku & 78.3 & 0.671 & 1.075 & 0.615 & $+0.280$ & $-1.36$ & $+0.040$ & $+2.12$ \\
\quad + Gemini & 79.4 & 0.686 & 1.063 & 0.631 & $+0.261$ & $-1.58$ & $+0.054$ & $+0.99$ \\
\quad + Qwen & 77.9 & 0.668 & 1.076 & 0.601 & $+0.280$ & $-1.93$ & $+0.034$ & $+0.96$ \\
\addlinespace
\bottomrule
\end{tabular}}
\end{table}

\paragraph{Paired comparisons.} Table~\ref{tab:llm_bisg_contrasts} compares each combination with BISG alone in the same bootstrap samples. The gains in accuracy and in the weighted estimator are distinguishable from zero for every LLM. The differences in BIRDiE's absolute bias are not, because BISG alone slightly overstates the registration gap and understates the age gap, while the combinations err in the other direction. The bias intervals themselves differ. On BISG alone both contain zero, and on every combined posterior both exclude it.

\paragraph{Possible explanations.} BIRDiE assumes that the outcome is independent of the name given race and geography. Table~\ref{tab:birdie_assumption} detects no association with the surname-based race probability, but the test has limited power. The combined posteriors carry information from the full name, and first names may carry birth-cohort information related to age within race. Such a relationship could violate BIRDiE's assumption and is consistent with the age pattern in Haiku's ``Unknown'' answers (Table~\ref{tab:missing_type}). The combination also produces overconfident probabilities by counting surname evidence twice. Our analysis does not separate these explanations. The weighted estimator's gains show that a combination can improve one estimator without establishing its suitability for another.

\begin{table}[!htbp]
\centering
\caption{Paired differences between BISG tract posteriors combined with an LLM's answer and BISG tract posteriors alone, Study 2. Positive accuracy differences favor the combination. For bias, the difference is in absolute bias, so negative values favor the combination. The last row compares the combination's accuracy with pyethnicity's. Brackets give 95\% stratified bootstrap intervals (500 replicates) in which $\Pr(L \mid R)$ is re-estimated in every replicate.}
\label{tab:llm_bisg_contrasts}
\scriptsize
\begin{tabular}{lccc}
\toprule
Difference & + Haiku & + Gemini & + Qwen \\
\midrule
Accuracy (points) & $+4.2$ [$+3.8$, $+4.7$] & $+5.4$ [$+4.8$, $+5.9$] & $+3.9$ [$+3.4$, $+4.4$] \\
Weighted, $|$bias$|$ party & $-0.027$ [$-0.034$, $-0.021$] & $-0.045$ [$-0.053$, $-0.038$] & $-0.027$ [$-0.033$, $-0.021$] \\
Weighted, $|$bias$|$ age & $-2.0$ [$-2.2$, $-1.7$] & $-1.7$ [$-2.0$, $-1.4$] & $-1.4$ [$-1.6$, $-1.1$] \\
BIRDiE, $|$bias$|$ party & $+0.020$ [$-0.022$, $+0.055$] & $+0.035$ [$-0.003$, $+0.069$] & $+0.014$ [$-0.026$, $+0.051$] \\
BIRDiE, $|$bias$|$ age & $+1.6$ [$-0.3$, $+2.5$] & $+0.5$ [$-1.3$, $+1.5$] & $+0.4$ [$-1.3$, $+1.5$] \\
Accuracy vs.\ pyethnicity (points) & $-0.2$ [$-0.8$, $+0.4$] & $+1.0$ [$+0.4$, $+1.5$] & $-0.5$ [$-1.1$, $-0.1$] \\
\bottomrule
\end{tabular}
\end{table}

\clearpage
\paragraph{Exploratory calibration check.} We replace the independence-based update with a ridge multinomial regression of self-reported race on the log tract-BISG probabilities, with or without indicators for the LLM's answer, missing-output types included. Probabilities are floored at $10^{-6}$ before taking logs, and predictors are standardized. We fix the penalty at $\lambda=0.001$ and use the same two halves as above, fitting on one and predicting the other. Neither outcome enters the model. Each of 500 race-stratified bootstrap samples refits the models, with copies of a voter kept in their original half. The calibration target is the stratified Study~2 sample.

LLM answers improve classification and probability scores relative to calibrated BISG alone (Tables~\ref{tab:llm_bisg_calibrated} and~\ref{tab:llm_bisg_calibrated_contrasts}). Gemini reaches 81.4\% accuracy against 77.4\% for calibrated BISG, a paired gain of 4.0 points (95\% interval 3.5 to 4.5). Adding each LLM reduces BIRDiE's absolute bias for the registration gap relative to calibrated BISG. For the age gap, absolute bias rises with Haiku and Gemini, and Qwen's interval includes zero. Every calibrated combination has greater absolute bias than original BISG for both gaps. At penalties of 0.0001 and 0.01, the LLM models still improve prediction over calibrated BISG and remain more biased than original BISG for both BIRDiE estimands. Calibration also increases BIRDiE's bias without an LLM, so the change cannot be attributed solely to LLM information.

\begin{table}[!htbp]
\centering
\caption{Exploratory held-out calibration and conditional LLM--BISG combination, Study 2. A ridge multinomial model predicts race from the log tract-BISG probabilities, with or without the LLM's output category, including missing outputs. Training uses the opposite half of the sample and excludes both outcomes. The fixed penalty is $\lambda=0.001$. Accuracy refers to the four-category sample; log loss and the multiclass Brier score use all five categories. Lower scores are better. Bias is for pooled BIRDiE and equals estimated minus true White--Black gap. Brackets give 95\% paired, race-stratified bootstrap intervals from 500 complete refits. Calibration targets the stratified sample.}
\label{tab:llm_bisg_calibrated}
\small
\begin{tabular}{lccccc}
\toprule
Posterior & Accuracy (\%) & Log loss & Brier & Party bias & Age bias \\
\midrule
Original BISG & 74.1 & 1.047 & 0.472 & \shortstack{$-0.020$\\{}[$-0.042$, $+0.001$]} & \shortstack{$-0.53$\\{}[$-1.60$, $+0.59$]} \\ [4pt]
\midrule
Calibrated BISG & 77.4 & 0.867 & 0.423 & \shortstack{$-0.103$\\{}[$-0.125$, $-0.076$]} & \shortstack{$+3.85$\\{}[$+2.02$, $+5.85$]} \\ [4pt]
+ Haiku & 80.0 & 0.803 & 0.393 & \shortstack{$-0.066$\\{}[$-0.086$, $-0.043$]} & \shortstack{$+6.28$\\{}[$+4.73$, $+7.85$]} \\ [4pt]
+ Gemini & 81.4 & 0.787 & 0.385 & \shortstack{$-0.062$\\{}[$-0.080$, $-0.039$]} & \shortstack{$+5.02$\\{}[$+3.55$, $+6.42$]} \\ [4pt]
+ Qwen & 79.9 & 0.810 & 0.397 & \shortstack{$-0.078$\\{}[$-0.097$, $-0.055$]} & \shortstack{$+4.57$\\{}[$+2.95$, $+6.07$]} \\ [4pt]
\bottomrule
\end{tabular}
\end{table}

\begin{table}[!htbp]
\centering
\caption{Paired contrasts for the exploratory conditional combination. Positive accuracy differences favor adding LLM answers; negative differences in log loss, Brier score, or absolute bias favor the combination. Each comparison uses the same 500 bootstrap samples and refits both calibration models. The first panel isolates the added predictive value of LLM outputs relative to calibrated BISG. The second compares inferential performance with the original BISG posteriors.}
\label{tab:llm_bisg_calibrated_contrasts}
\small
\begin{tabular}{lccc}
\toprule
Difference & + Haiku & + Gemini & + Qwen \\
\midrule
\multicolumn{4}{l}{\textit{Relative to calibrated BISG alone}} \\ [3pt]
Accuracy (points) & \shortstack{$+2.7$\\{}[$+2.3$, $+3.2$]} & \shortstack{$+4.0$\\{}[$+3.5$, $+4.5$]} & \shortstack{$+2.5$\\{}[$+2.2$, $+3.1$]} \\ [4pt]
Log loss & \shortstack{$-0.063$\\{}[$-0.068$, $-0.056$]} & \shortstack{$-0.080$\\{}[$-0.085$, $-0.072$]} & \shortstack{$-0.057$\\{}[$-0.061$, $-0.051$]} \\ [4pt]
Brier score & \shortstack{$-0.030$\\{}[$-0.033$, $-0.027$]} & \shortstack{$-0.039$\\{}[$-0.042$, $-0.035$]} & \shortstack{$-0.026$\\{}[$-0.029$, $-0.023$]} \\ [4pt]
BIRDiE, $|$bias$|$ party & \shortstack{$-0.037$\\{}[$-0.050$, $-0.026$]} & \shortstack{$-0.041$\\{}[$-0.053$, $-0.030$]} & \shortstack{$-0.025$\\{}[$-0.035$, $-0.016$]} \\ [4pt]
BIRDiE, $|$bias$|$ age & \shortstack{$+2.43$\\{}[$+1.49$, $+3.33$]} & \shortstack{$+1.17$\\{}[$+0.18$, $+2.09$]} & \shortstack{$+0.72$\\{}[$-0.15$, $+1.40$]} \\ [4pt]
\midrule
\multicolumn{4}{l}{\textit{Relative to original BISG}} \\ [3pt]
BIRDiE, $|$bias$|$ party & \shortstack{$+0.046$\\{}[$+0.029$, $+0.059$]} & \shortstack{$+0.042$\\{}[$+0.026$, $+0.054$]} & \shortstack{$+0.058$\\{}[$+0.041$, $+0.070$]} \\ [4pt]
BIRDiE, $|$bias$|$ age & \shortstack{$+5.76$\\{}[$+3.28$, $+7.51$]} & \shortstack{$+4.50$\\{}[$+1.94$, $+6.23$]} & \shortstack{$+4.05$\\{}[$+1.51$, $+5.87$]} \\ [4pt]
\bottomrule
\end{tabular}
\end{table}

\newpage
\section{Downstream Bias for the North Carolina Voter Population}
\label{app:popdownstream}

The estimands in Section~\ref{sec:error_propagation} are defined on the stratified sample. Table~\ref{tab:downstream_popweighted} reweights every voter by the ratio of the group's share in the full North Carolina voter file (69.2\% White, 21.5\% Black, 4.4\% Hispanic, 1.9\% Asian, and 3.0\% other) to its share in the sample, and recomputes the estimands and every estimator. The weights enter each estimator voter by voter, including the BIRDiE fits. Reweighting the estimated group means after estimation would leave every estimate of a gap unchanged. Reweighting voters changes the mix of true groups among the voters each tool labels White or Black, which is what moves the estimators. The true gaps do not change, because the weights are constant within self-reported race, but the Black share does. BISG and pyethnicity posteriors embed population priors, so the reweighted sample, whose composition matches the voter file, is the setting in which those priors are closest to correct.

\begin{table}[!htbp]
\centering
\caption{Bias in downstream estimands with every voter weighted to the race composition of the full North Carolina voter file. The weights enter each estimator observation by observation, so they change the mix of true groups inside each predicted group. The population estimands are $\pi_B = 0.215$, $\Delta_{\text{party}} = -0.529$, and $\Delta_{\text{age}} = 4.29$ years. The two gaps equal the sample gaps, because weights that are constant within self-reported race leave the true group means unchanged. Brackets give 95\% paired bootstrap intervals for the bias, as in Table~\ref{tab:downstream_bias}.}
\label{tab:downstream_popweighted}
\footnotesize
\begin{tabular}{lccc}
\toprule
Estimator & Bias($\hat{\pi}_B$) & Bias($\hat{\Delta}_{\text{party}}$) & Bias($\hat{\Delta}_{\text{age}}$, years) \\
\midrule
\multicolumn{4}{l}{\textit{Hard-label (argmax) estimators}} \\
\quad WRU surname & \shortstack{$-0.161$\\{}[$-0.164$, $-0.157$]} & \shortstack{$+0.292$\\{}[$+0.251$, $+0.330$]} & \shortstack{$-2.29$\\{}[$-3.77$, $-0.94$]} \\
\quad WRU +county & \shortstack{$+0.003$\\{}[$-0.004$, $+0.010$]} & \shortstack{$+0.274$\\{}[$+0.249$, $+0.294$]} & \shortstack{$-2.00$\\{}[$-3.01$, $-1.00$]} \\
\quad WRU +tract & \shortstack{$+0.000$\\{}[$-0.006$, $+0.007$]} & \shortstack{$+0.217$\\{}[$+0.195$, $+0.237$]} & \shortstack{$-1.53$\\{}[$-2.53$, $-0.66$]} \\
\quad pyethnicity & \shortstack{$-0.003$\\{}[$-0.010$, $+0.004$]} & \shortstack{$+0.221$\\{}[$+0.201$, $+0.240$]} & \shortstack{$-1.45$\\{}[$-2.40$, $-0.47$]} \\
\quad Claude Haiku 4.5 & \shortstack{$-0.097$\\{}[$-0.104$, $-0.090$]} & \shortstack{$+0.134$\\{}[$+0.094$, $+0.166$]} & \shortstack{$+7.39$\\{}[$+5.90$, $+8.85$]} \\
\quad Gemini 2.5 Flash & \shortstack{$-0.061$\\{}[$-0.066$, $-0.055$]} & \shortstack{$+0.235$\\{}[$+0.210$, $+0.257$]} & \shortstack{$+0.20$\\{}[$-0.81$, $+1.26$]} \\
\quad Qwen 2.5 7B (local) & \shortstack{$+0.007$\\{}[$-0.001$, $+0.014$]} & \shortstack{$+0.333$\\{}[$+0.310$, $+0.357$]} & \shortstack{$-0.90$\\{}[$-1.95$, $+0.36$]} \\
\quad Hybrid Haiku$\rightarrow$pyeth. & \shortstack{$-0.014$\\{}[$-0.020$, $-0.008$]} & \shortstack{$+0.221$\\{}[$+0.201$, $+0.241$]} & \shortstack{$-0.93$\\{}[$-1.87$, $+0.02$]} \\
\addlinespace
\multicolumn{4}{l}{\textit{Posterior-weighting estimators}} \\
\quad WRU surname & \shortstack{$-0.035$\\{}[$-0.038$, $-0.032$]} & \shortstack{$+0.433$\\{}[$+0.417$, $+0.449$]} & \shortstack{$-3.38$\\{}[$-4.14$, $-2.60$]} \\
\quad WRU +county & \shortstack{$+0.029$\\{}[$+0.025$, $+0.033$]} & \shortstack{$+0.368$\\{}[$+0.352$, $+0.384$]} & \shortstack{$-2.72$\\{}[$-3.45$, $-1.98$]} \\
\quad BISG tract chain & \shortstack{$+0.020$\\{}[$+0.016$, $+0.025$]} & \shortstack{$+0.307$\\{}[$+0.292$, $+0.322$]} & \shortstack{$-2.55$\\{}[$-3.24$, $-1.89$]} \\
\quad pyethnicity & \shortstack{$+0.056$\\{}[$+0.052$, $+0.060$]} & \shortstack{$+0.324$\\{}[$+0.309$, $+0.338$]} & \shortstack{$-2.00$\\{}[$-2.64$, $-1.30$]} \\
\addlinespace
\multicolumn{4}{l}{\textit{Mixed LLM + posterior pipelines}} \\
\quad Haiku + pyeth.\ posterior & \shortstack{$+0.032$\\{}[$+0.028$, $+0.037$]} & \shortstack{$+0.310$\\{}[$+0.295$, $+0.324$]} & \shortstack{$-1.31$\\{}[$-1.98$, $-0.59$]} \\
\quad Haiku + BISG posterior & \shortstack{$+0.004$\\{}[$-0.000$, $+0.008$]} & \shortstack{$+0.277$\\{}[$+0.261$, $+0.292$]} & \shortstack{$-1.08$\\{}[$-1.74$, $-0.37$]} \\
\addlinespace
\multicolumn{4}{l}{\textit{BIRDiE estimators}} \\
\quad BIRDiE, WRU surname & \shortstack{$-0.037$\\{}[$-0.040$, $-0.033$]} & \shortstack{$-0.044$\\{}[$-0.094$, $+0.007$]} & \shortstack{$+1.09$\\{}[$-1.64$, $+3.46$]} \\
\quad BIRDiE, WRU +county & \shortstack{$+0.026$\\{}[$+0.021$, $+0.031$]} & \shortstack{$+0.006$\\{}[$-0.027$, $+0.042$]} & \shortstack{$+1.05$\\{}[$-0.66$, $+2.62$]} \\
\quad BIRDiE, BISG tract chain & \shortstack{$+0.016$\\{}[$+0.011$, $+0.020$]} & \shortstack{$-0.024$\\{}[$-0.051$, $+0.004$]} & \shortstack{$+0.27$\\{}[$-1.03$, $+1.48$]} \\
\addlinespace
\multicolumn{4}{l}{\textit{Reference}} \\
\quad Oracle (true labels) & 0 (by construction) & 0 (by construction) & 0 (by construction) \\
\addlinespace
\bottomrule
\end{tabular}
\end{table}

\newpage
\section{Temporal Holdout Test for Training-Data Contamination}
\label{app:temporal}

The North Carolina voter file is public, so the LLMs may have seen the self-reported race of sampled voters during training. Voters who registered after a model's training cutoff cannot appear in any snapshot it was trained on. We use each model's documented cutoff (SI~\ref{app:models}) and drop the month after it as a buffer. If performance reflected memorized records, accuracy should be lower for voters registered after the cutoff. Table~\ref{tab:temporal} shows no such drop. For Haiku, whose training data run to July 2025, the 675 later registrants are labeled correctly 81.5\% of the time against 82.6\% for earlier registrants, and the interval excludes a drop larger than 5.0 points. Gemini and Qwen are slightly more accurate on later registrants, and the difference loses significance once race is controlled.

\begin{table}[!htbp]
\centering
\caption{Temporal holdout test, Study 2. Voters who registered after a model's documented training cutoff (with a one-month buffer dropped) cannot appear in any voter-file snapshot the model was trained on. Differences are post minus pre, with 95\% stratified bootstrap intervals. The logit coefficient is from a regression of correctness on a post-cutoff indicator and self-reported race among labeled voters.}
\label{tab:temporal}
\small
\begin{tabular}{llccccc}
\toprule
Model & Cutoff & $N$ post & Cond.\ acc.\ pre & Cond.\ acc.\ post & Difference (pp) & Logit (SE) \\
\midrule
Claude Haiku 4.5 & Jul 2025 &   675 & 82.6 & 81.5 & $-1.1$ [$-5.0$, $+2.7$] & $-0.12$ (0.18) \\
Gemini 2.5 Flash & Jan 2025 & 1,259 & 71.1 & 74.9 & $+3.8$ [$+1.3$, $+6.3$] & $0.16$ (0.09) \\
Qwen 2.5 7B & Sep 2024 & 1,767 & 66.0 & 70.7 & $+4.8$ [$+2.4$, $+6.9$] & $0.12$ (0.07) \\
\bottomrule
\end{tabular}
\end{table}

This test addresses memorization of voter records. It does not address the models' general exposure to associations between names and race, which is what makes them useful classifiers. A validation on a voter file that is not publicly available would address the remaining concern.
}

\ifdefined\preprintversion\else
\documentclass[12pt]{article}
\fi

\usepackage{booktabs}
\usepackage{graphicx}
\usepackage{amsmath}
\usepackage{amssymb}
\usepackage[margin=1in]{geometry}
\usepackage{caption}
\usepackage{xr-hyper}
\usepackage{enumitem}
\usepackage[authoryear,round]{natbib}
\usepackage[T1]{fontenc}
\usepackage{tabularx}
\usepackage{threeparttable}
\usepackage{xcolor}
\usepackage{setspace}
\usepackage{float}
\usepackage{hyperref}
\ifdefined\preprintversion\else
\fi

\newif\ifanon
\ifdefined\preprintversion\anonfalse\else\anontrue\fi

\hypersetup{
  colorlinks = true,
  linkcolor  = blue!70!black,
  citecolor  = blue!70!black,
  urlcolor   = blue!70!black,
  pdftitle   = {Who Gets Counted?},
  pdfauthor  = {},
}

\title{\textbf{\Large Who Gets Counted? Large Language Models, Informative Refusal, and Downstream Bias in Name-Based Demographic Classification}}
\ifanon
\author{}
\else
\ifdefined\preprintauthors
\author{\preprintauthors}
\else
\author{Christian Pipal, Morgan Wack, and Eva-Maria Vogel}
\fi
\fi
\date{}

\begin{document}
\maketitle

\begin{abstract}
\noindent
Researchers routinely infer gender, race, and ethnicity from names, yet the standard tools used for this task misclassify the groups that diversity research most needs to measure. Large language models (LLMs) have been proposed as a remedy because they label diverse names accurately. We ask whether that accuracy carries over to the estimates researchers report. Across 10,000 Chinese names, 19,998 North Carolina voters with self-reported race, and 300 political science authors with human reference codes, we compare three LLMs with seven traditional tools and link the voter sample to party registration and age. Selective labeling explains part of the LLMs' accuracy advantage. Claude Haiku labels half of the voters, and a traditional tool limited to its most confident half is more accurate. Hard labels from every tool understate the White--Black gap in Democratic registration by 36\% to 65\%, while BIRDiE estimation on BISG posteriors comes within 0.03 of it. Haiku's missing labels select on age, inflating the age gap by six years. Combined with BISG, LLM answers sharpen individual labels without a demonstrated gain in BIRDiE accuracy. Classifiers should be judged by the estimates they produce, including the consequences of missing labels.
\end{abstract}

\vspace{0.5em}
\noindent\textbf{Keywords:} demographic classification, large language models, measurement error, informative missingness, computational social science

\ifdefined\preprintversion
\onehalfspacing
\else
\setcounter{page}{1}
\doublespacing
\fi
\setlength{\parskip}{0pt}
\setlength{\parindent}{1.5em}

\section{Introduction}
\label{sec:intro}

Studies of representation in citation networks, hiring, legislatures, and public office routinely infer the race and gender of individuals from their names, while computational tools have been used to scale these practices to millions of records \citep{dion2018gendered, dworkin2020extent, bertrand2004emily, imai2016improving, gross2024measuring}. Yet the tools that make this possible fail unevenly. Threshold-based Bayesian Improved Surname Geocoding (BISG), the standard method for race, undercounts Black individuals by about 19\% and overcounts White individuals by 16\% \citep{kozlowski2022intersectional}. In a survey of 19,924 scholars who reported their own identities, the best-performing gender tool misgendered 43\% of Chinese women \citep{lockhart2023namebased}. Classification errors often concentrate among the groups that diversity research sets out to count.

Large language models (LLMs) appear to address some of these challenges. Given only a name, they have been shown to assign the correct self-reported race to between 75\% and 85\% of people in recent benchmarks, while their errors fall less heavily on non-White names than those of surname dictionaries \citep{alnuaimi2024enriching, dasanaike2026llm}. For gender, leading LLMs have the lowest error rates of the tools tested \citep{dominguezdiaz2024gender}. These benchmarks rank classifiers by their accuracy among the names each tool labels, which remains a focus for researchers when choosing a tool.

Accuracy across labeled names leaves out two things that decide whether a classifier improves applied work. One is which names each tool declines to label. Dropping unlabeled cases biases estimates whenever the reason a case is missing is related to the outcome \citep{king2001analyzing, knox2020administrative}. This is particularly relevant as safety-tuned LLMs often refuse requests to classify race. As a result, where refusals fall on hard names or on particular groups, high accuracy on the labeled names says little about the full sample. The other is what the labels do to the estimand. Researchers use inferred categories to estimate group shares and group differences, and misclassification can pull group differences toward zero even when accuracy is high \citep{blackwell2017unified, chen2019fairness, argyle2024misclassification}. Existing benchmarks that do not track refusal and downstream bias thus cannot tell us whether LLMs improve the estimates researchers publish.

We expect refusal to work as a selection mechanism. An LLM that answers when a name is easy and declines when it is ambiguous will post high accuracy by labeling names that every tool gets right. If the ambiguity of a name is related to the outcome under study, estimates computed on the labeled subset are biased even when every label is correct.

We test this across three purposefully selected validation datasets. Two hosted LLMs, a locally run open-weight LLM, and seven traditional tools classify 10,000 Chinese Pinyin names with known gender frequencies among their bearers (Study 1), 19,998 North Carolina voters with self-reported race (Study 2), and 300 cited political science authors whom a research assistant coded from public profiles (Study 3). Study 1 tests classification under name ambiguity, and Study 3 tests performance in an applied sample with perceived-identity codes. For Study 2 we link each sampled voter back to the voter file, which adds party registration and age to true race and every tool's prediction. These linked data let us compute exactly how much each classification pipeline biases three estimands and compare the hard-label practice of most applied work with the weighting estimator analyzed by \citet{chen2019fairness} and the BIRDiE estimator of \citet{mccartan2024birdie}.

Selective labeling accounts for Haiku's apparent advantage over pyethnicity. Claude Haiku is correct on 89.3\% of the voters it labels but labels only half of them. On the same voters, pyethnicity, a neural classifier trained on voter files, is correct 90.0\% of the time, and restricted to its most confident half it reaches 94.1\%. For the registration gap, the choice of estimator matters more than the choice of classifier. Hard labels from every tool understate the White--Black gap in Democratic registration by 36\% to 65\%, and posterior weighting by 58\% to 81\%, while BIRDiE applied to BISG posteriors misses the true gap of $-0.529$ by 0.012. Informative missingness also produces selection bias that a correction based on a single confusion matrix cannot remove. Haiku's ``Unknown'' answers fall disproportionately on older Black voters, and its hard labels overstate the White--Black age gap by 6.0 years against a true gap of 4.3 years. Filling its missing labels with a full-coverage tool removes the selection part of this bias, but the misclassification part remains.

Our comparison links classification performance to downstream bias by separating misclassification from selection. Combined with BISG posteriors, LLM answers, missing outputs included, raise individual accuracy by 3.9 to 5.4 points, and with Gemini's answers the combination is the most accurate classifier in our study. This gain does not establish that the combined posteriors improve estimates of gaps, which also depend on calibration and estimator assumptions. Where no validated estimator of downstream disparities exists, as for gender and non-U.S.\ names, LLMs label names traditional tools miss, but their labels still have to be validated against the estimand before they are used.

\section{Name-Based Classification and Its Blind Spots}
\label{sec:background}

Table~\ref{tab:tool_families} summarizes the tools we compare. \textit{Census-based tools} (WRU and predictrace) combine U.S.\ Census surname distributions with optional geographic priors \citep{imai2016improving, imai2022fbisg}. \textit{LSTM-based tools} (ethnicolr, rethnicity, and pyethnicity) read names as character sequences and are trained on voter registration data from one to fifty states \citep{chintalapati2018predicting, xie2022rethnicity, chintalapati2018pyethnicity}. For gender, Genderize.io aggregates cross-national first-name frequencies, and the \texttt{gender} R package draws on U.S.\ Social Security records \citep{mullen2015gender}. All of these tools learn from populations that are well represented in U.S.\ and Western administrative records, and SI~\ref{app:tools} describes each of them.

\begin{table}[htbp]
\centering
\caption{Classification tool families compared in this paper}
\label{tab:tool_families}
\footnotesize
\resizebox{\textwidth}{!}{%
\begin{tabular}{llll}
\toprule
\textbf{Family} & \textbf{Tools} & \textbf{Input} & \textbf{Training data} \\
\midrule
\multicolumn{4}{l}{\textit{Race and ethnicity}} \\
Census-based & WRU, predictrace & Surname $\pm$ geography & U.S.\ Census \\
LSTM-based & ethnicolr, rethnicity, pyethnicity & Full name & Voter files \\
LLM, hosted & Claude Haiku 4.5, Gemini 2.5 Flash & Full name + country & Web text \\
LLM, local & Qwen 2.5 7B Instruct & Full name + country & Web text \\
\midrule
\multicolumn{4}{l}{\textit{Gender}} \\
Frequency-based & Genderize.io & First name & Cross-national frequencies \\
Dictionary-based & \texttt{gender} R package & First name & U.S.\ SSA records \\
LLM & Haiku, Gemini, Qwen & Full name + country & Web text \\
\bottomrule
\end{tabular}}
\end{table}

The failures of these tools are structured. In the survey by \citet{lockhart2023namebased}, gender misclassification was 3.5 times more common for women than for men, every tool misclassified all 139 nonbinary respondents, and race misrecognition was highest for Black and Filipino scholars. These concentrated errors move estimates in a direction set by the pattern of errors and by the estimand \citep{blackwell2017unified}.

\subsection{LLMs as Classifiers}

LLMs now serve as measurement instruments across the social sciences. They annotate text more accurately than crowd workers \citep{gilardi2023chatgpt}, place political texts on ideological scales \citep{ornstein2025positioning}, and code concepts from unstructured documents \citep{halterman2025codebook, pangakis2023automated}. A single model covers naming traditions that no one training set spans, which is why benchmarks find LLMs competitive with or better than purpose-built name classifiers, with the largest gains for names the tools handle worst \citep{alnuaimi2024enriching, dasanaike2026llm, dominguezdiaz2024gender}.

The benchmarks share a scoring rule that sets aside two features of LLM classification. LLMs refuse, and conditional accuracy excludes refusals from the denominator. And the labels are an input to an estimate, so the relevant error is the error in the estimate, which accuracy does not measure.

\subsection{From Labels to Estimates}
\label{sec:theory}

Two mechanisms connect a classifier's errors to bias in the estimates built on its labels. The first is misclassification. When researchers treat a tool's most likely category as the true category, each predicted group mixes members of several true groups. The mean outcome in a predicted group then moves toward the means of the groups it absorbs, and group differences shrink toward zero \citep{blackwell2017unified, argyle2024misclassification}. Using a tool's posterior probabilities as weights does not eliminate this problem. \citet{chen2019fairness} show that the weighted estimator also attenuates group differences when posteriors overlap and the outcome is independent of the name given race, and that the hard-label estimator has a bias with no simple general form. With more than two groups and with selective labeling, hard-label bias can push a gap in either direction. BIRDiE \citep{mccartan2024birdie} models the outcome and race jointly, so that the outcome itself informs each individual's race probabilities. It requires probabilistic race predictions and assumes that the outcome is independent of the name given race and geography. We therefore expect weighted estimates to understate group differences, hard-label estimates to be biased in a direction set by the pattern of errors, and BIRDiE to come close to the truth when its assumption holds.

The second mechanism is selection. A classifier that declines uncertain cases is a classifier with a reject option \citep{chow1970optimum, herbei2006classification}. Rejecting uncertain cases raises accuracy on the cases that remain, and if rejection is unrelated to the outcome, the only cost is a smaller sample. If rejection is related to the outcome, estimates on the retained cases are biased \citep{elyaniv2010foundations, geifman2017selective}. The same concern arises whenever the process that produces missing data is related to the variable of interest \citep{mitra2023learning}, and health researchers have documented it for race records, where missing self-reports bias estimated disparities \citep{yee2022implications, krobath2025nonrandom}. LLM refusals come from safety training and from the model's uncertainty about a name, so we expect them to fall disproportionately on ambiguous names and on some groups. This yields two expectations. The accuracy advantage of LLMs should be concentrated among names that other tools also classify correctly, and hard-label LLM estimates should carry a selection component that full-coverage tools lack.

\section{Data and Method}
\label{sec:method}

All tools classify the same names in each study. As a result, every comparison is within names. Study~1 tests gender classification on 10,000 Chinese Pinyin names with known gender frequencies. Study~2 tests race and ethnicity classification on 19,998 North Carolina voters with self-reported race and adds real outcomes from the voter file. Study~3 tests both tasks on 300 cited political science authors whom a research assistant coded from public profiles without seeing any algorithmic prediction.

Three metrics recur. \textit{Coverage} is the share of names for which a tool returns a usable label. A name can go unlabeled because its surname is missing from a Census dictionary, because an API returns no prediction, or because an LLM refuses, and SI~\ref{app:coverage} describes the mechanism for each tool. \textit{Conditional accuracy} is the share of correct labels among labeled names. \textit{Total accuracy} is the product of the two, the probability that a randomly drawn name receives the correct label. We report all three because a tool can score high on conditional accuracy while labeling few names. An LLM leaves a name unlabeled in three ways. It can refuse outright, answer ``Unknown'', or give a category outside the four we score, mainly Middle Eastern (SI~\ref{app:missing}). We use ``refusal'' for all three when the distinction does not matter.

The hosted LLMs are Claude Haiku 4.5 \citep{anthropic2025claude45haiku} and Gemini 2.5 Flash \citep{geminiteam2025gemini25}. Both received the same short instruction, which asked only for a category, together with the name and a country code (``US'' for voters, ``CN'' for Study~1, and the country of affiliation for Study~3), at temperature zero. Repeated runs on 200 names reproduced 100\% of Haiku's labels and 99.5\% of Gemini's (SI~\ref{app:determinism}). Hosted models can still be updated or retired, which limits long-run reproducibility \citep{barrie2025replication}. We therefore also ran a locally hosted open-weight model, Qwen 2.5 7B Instruct at the pinned quantization \texttt{qwen2.5:7b-instruct-q4\_K\_M}, with the same instruction, temperature zero, and a fixed seed. A full second run on 2,000 Study~2 names reproduced all of its labels. The hosted models cost about \$28 in API fees for roughly 40,000 names per model. The local model runs offline at no marginal cost. We chose the instruction from four variants tested on 200 names per task. Every variant that lowered refusal rates also lowered conditional accuracy, and SI~\ref{app:prompts} gives the exact wording of all variants.

\section{Study 1: Gender Classification of Chinese Names}
\label{sec:study1}

Chinese names are the hardest case for gender classification. Misgendering rates for Chinese women range from 43\% to 76\% across tools, the highest for any group in the literature \citep{lockhart2023namebased}, because many Pinyin names are used by both men and women and standard tools have little training data for them \citep{sebo2021performance}. If LLMs help anywhere, it should be here.

We sampled 10,000 Pinyin given names from the dataset of \citet{shi2025chinese}, which records the gender of about 30 million individual shareholders of companies registered in China up to 2018 and aggregates their given names into 96,797 Pinyin forms.\footnote{Available at \url{https://doi.org/10.7910/DVN/UAARYY}.} The gender shares are frequencies among these shareholders, not in the Chinese population. We kept the 54,897 Pinyin names with at least ten bearers and drew 2,500 names from each of four strata of the female share (above 80\%, 50\% to 80\%, 20\% to 50\%, and below 20\%). No sampled name has an even split. The sample oversamples ambiguous names by design and does not represent the distribution of Chinese individuals.

We score tools against two targets. \textit{Majority agreement} is the share of labeled names whose label is the name's majority gender, a target a perfect classifier could meet for every name. \textit{Individual accuracy} is the expected accuracy for a person drawn from the bearers of a labeled name, so that a female label earns the name's female share. Its ceiling is the mean majority share, which a classifier that always picks the majority gender would reach. Across the 10,000 names this ceiling is 79.7\%, and because tools label different names, Table~\ref{tab:study1} reports it for each tool's labeled names (SI~\ref{app:ceiling}). We queried Genderize.io without its optional country parameter. Setting it to China raises conditional accuracy by 3.3 percentage points but lowers coverage from 79\% to 50\% (SI~\ref{app:genderize}).

\begin{table}[htbp]
\centering
\caption{Gender classification of 10,000 Chinese Pinyin names (Study 1)}
\label{tab:study1}
\begin{threeparttable}
\footnotesize
\begin{tabular}{lccccc}
\toprule
 & & \textbf{Majority agreement} & \textbf{Total} & \textbf{Individual} & \textbf{Ceiling on} \\
\textbf{Tool} & \textbf{Coverage} & \textbf{[95\% CI]} & \textbf{agreement} & \textbf{accuracy} & \textbf{labeled names} \\
\midrule
Genderize.io & 79.3 & 70.4 [69.4, 71.3] & 55.8 & 64.8 & 79.0 \\
\texttt{gender} R package & 3.8 & 56.5 [51.5, 61.6] & 2.1 & 55.7 & 81.0 \\
\midrule
Gemini 2.5 Flash & 28.4 & 87.7 [86.4, 88.9] & 24.9 & 78.7 & 84.0 \\
Claude Haiku 4.5 & 8.2 & 85.1 [82.7, 87.7] & 7.0 & 78.9 & 85.6 \\
Qwen 2.5 7B (local) & 7.6 & 81.7 [79.0, 84.3] & 6.2 & 75.3 & 84.2 \\
\bottomrule
\end{tabular}
\begin{tablenotes}[flushleft]\small
\item Percent, averaged over names. Total agreement counts unlabeled names as errors. Individual accuracy and its ceiling are defined in the text. Across all names the ceiling is 79.7\%. Intervals from 2,000 bootstrap replications. Person-weighted versions are in SI~\ref{app:ceiling}.
\end{tablenotes}
\end{threeparttable}
\end{table}

Genderize.io labels 79.3\% of names and agrees with the majority gender on 70.4\% of them (Table~\ref{tab:study1}). The LLMs are far more selective. Haiku returns ``Unknown'' for 91.8\% of names and Gemini for 71.6\%, but they agree with the majority gender on 85.1\% and 87.7\% of the names they label. Part of this advantage reflects which names they label, since the individual-level ceiling on their labeled names is 85.6\% and 84.0\%, against 79.0\% for Genderize.io. Scored on the same names, the advantage shrinks or disappears. On the 2,308 names both Gemini and Genderize.io label, they agree with the majority gender on 87.7\% and 80.9\%, and on the 735 names both Haiku and Genderize.io label, both agree on 85.4\%. The local Qwen model behaves like Haiku, labeling 7.6\% at 81.7\%. Among the most ambiguous names every tool's individual accuracy lies between 50.9\% and 52.0\%, against a ceiling of 55.5\%, while among names whose majority share exceeds 80\% Haiku and Gemini agree with the majority gender on 94.4\% and 95.7\%, against 80.5\% for Genderize.io (SI~\ref{app:study1ambiguity}). The LLMs behave like classifiers with an internal confidence threshold.

For female names, which tools misclassify most often \citep{lockhart2023namebased}, the LLMs trade recall for precision. Haiku labels 10.4\% of majority-female names female and Gemini 40.8\%, against 61.6\% for Genderize.io, but a name they label female is majority female 91.7\% and 87.9\% of the time, against 69.2\%. On the 2,069 names Genderize.io cannot label, Gemini labels 25.9\% and agrees with the majority gender on 87.7\%, while Haiku labels 4.2\% at 82.6\%. For Chinese names, LLM accuracy holds in the coverage gap, but the LLMs fill only a small part of it.

\section{Study 2: Race and Ethnicity of North Carolina Voters}
\label{sec:study2}

Race classification relies on Census surname distributions and voter records, and the groups those data represent worst are the groups tools misclassify most \citep{lockhart2023namebased}. North Carolina is one of few states whose voter file records self-reported race, which lets us score every tool against the voter's own answer.

The voter file (snapshot of March 2026) contains 9.1 million records.\footnote{\url{https://www.ncsbe.gov/results-data/voter-registration-data}} We drew a stratified sample of 19,998 voters, 5,000 each who report White, Black, and Hispanic identity, 2,998 who report Asian identity, and 2,000 who report another category (60\% ``Other'' race, 26\% American Indian, and 14\% multiracial). The sample therefore differs from the composition of North Carolina voters and from the Census populations whose composition enters the priors of BISG and other probabilistic tools. Our estimates describe the sample. SI~\ref{app:popweight} reports classification accuracy under U.S.\ Census population weights, and SI~\ref{app:popdownstream} reweights downstream estimates to the North Carolina voter-file composition. The Census-based tools are WRU with surname only, with county priors, and with census-tract priors, plus predictrace. For the tract version we geocoded each voter's residential address with the Census geocoder. Tract priors are the geography that BISG practice recommends \citep{imai2016improving}, and 79\% of the sample geocoded to a tract, with county priors used for the rest.\footnote{We do not report WRU's first-name specification, because in our implementation it returned predictions identical to the surname-only model.} The LSTM tools are rethnicity, ethnicolr, and pyethnicity, and the LLMs are Haiku, Gemini, and Qwen. The tools receive different information. Census tools see the surname and optionally the address, LSTM tools see the full name, and LLMs see the full name and the country. An ablation study that varies what the LLMs see is reported in SI~\ref{app:ablation}. We also build four hybrid pipelines that use an LLM's label when one exists and fall back to a traditional tool otherwise, so that the complementary workflow sits in the same comparison as its parts.

\begin{table}[!htbp]
\centering
\caption{Race and ethnicity classification of North Carolina voters (Study 2, $N = 17{,}998$ voters who report White, Black, Hispanic, or Asian identity). Coverage is the share of voters receiving a usable label. Conditional accuracy is the share correct among labeled voters, and total accuracy is their product. Hybrid pipelines use the LLM's label when one exists and the second tool's label otherwise. Brackets give 95\% intervals from 2,000 stratified bootstrap replications.}
\label{tab:study2_overall_v2}
\small
\begin{tabular}{lccc}
\toprule
Tool & Coverage & Cond. accuracy & Total accuracy \\
\midrule
pyethnicity & 100.0 & 78.5 [77.9, 79.0] & 78.5 [77.9, 79.0] \\
WRU (+tract) & 99.6 & 74.1 [73.5, 74.7] & 73.8 [73.2, 74.5] \\
rethnicity & 100.0 & 73.7 [73.0, 74.3] & 73.7 [73.0, 74.3] \\
WRU (+county) & 99.5 & 70.7 [70.1, 71.3] & 70.4 [69.8, 71.0] \\
ethnicolr & 100.0 & 67.9 [67.3, 68.5] & 67.9 [67.3, 68.5] \\
WRU (surname) & 99.9 & 63.5 [62.9, 64.0] & 63.4 [62.9, 64.0] \\
predictrace & 85.3 & 63.5 [62.9, 64.1] & 54.1 [53.6, 54.7] \\
\midrule
Gemini 2.5 Flash & 90.7 & 78.6 [78.0, 79.2] & 71.2 [70.6, 71.9] \\
Qwen 2.5 7B (local) & 95.6 & 73.5 [72.8, 74.1] & 70.3 [69.6, 70.9] \\
Claude Haiku 4.5 & 50.5 & 89.3 [88.6, 89.9] & 45.1 [44.4, 45.7] \\
\midrule
Hybrid: Haiku $\rightarrow$ pyethnicity & 100.0 & 78.1 [77.5, 78.6] & 78.1 [77.5, 78.6] \\
Hybrid: Haiku $\rightarrow$ WRU (+tract) & 99.8 & 77.4 [76.8, 77.9] & 77.2 [76.6, 77.8] \\
Hybrid: Gemini $\rightarrow$ pyethnicity & 100.0 & 76.7 [76.1, 77.3] & 76.7 [76.1, 77.3] \\
Hybrid: Haiku $\rightarrow$ WRU (+county) & 99.7 & 74.9 [74.3, 75.5] & 74.7 [74.1, 75.3] \\
\bottomrule
\end{tabular}
\begin{minipage}{0.95\textwidth}\vspace{4pt}\footnotesize
The 2,000 sampled voters who report another category are excluded because no tool can return that category. SI~\ref{app:other} reports how the tools label them.
\end{minipage}
\end{table}

\begin{figure}[htbp]
\centering
\includegraphics[width=\textwidth]{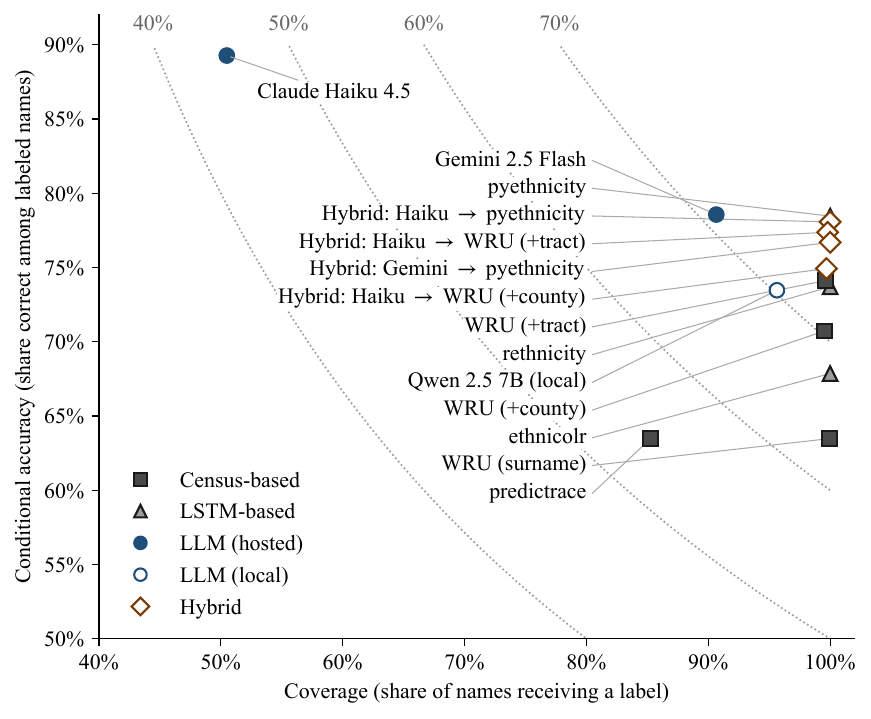}
\caption{Coverage and conditional accuracy of every tool and hybrid pipeline, Study 2 ($N = 17{,}998$). Dotted curves mark constant total accuracy. Haiku reaches the highest conditional accuracy by labeling about half of the voters, and the hybrids that fill its refusals land beside pyethnicity.}
\label{fig:coverage_frontier}
\end{figure}

Table~\ref{tab:study2_overall_v2} scores the tools on the 17,998 voters who report one of the four categories the tools can return (SI~\ref{app:other} reports how the tools label the others). Surname-only tools reach 63.5\%, the ceiling set by Census surname distributions. County priors raise WRU to 70.7\% and tract priors to 74.1\%, and pyethnicity reaches 78.5\%. Haiku has the highest conditional accuracy at 89.3\% but labels only 50.5\% of voters, so its total accuracy of 45.1\% is the lowest in the table. Gemini labels 90.7\% at 78.6\%, and Qwen 95.6\% at 73.5\%. Haiku's missing labels are 21 points of explicit refusals and 29 points of ``Unknown'' answers (SI~\ref{app:missing}). The hybrids restore nearly full coverage, and the best of them, Haiku with a pyethnicity fallback, reaches 78.1\% against 78.5\% for pyethnicity alone. Figure~\ref{fig:coverage_frontier} places every pipeline on the coverage and accuracy plane.

Conditional accuracy scores each tool on different voters, so it mixes the quality of the labels with the choice of which voters to label. We compare tools on common voters and at matched coverage (SI~\ref{app:paired}). On the 9,088 voters that both Haiku and pyethnicity label, pyethnicity is correct on 90.0\% and Haiku on 89.3\%. If pyethnicity labels only its most confident voters, as many as Haiku labels, it is correct on 94.1\%, against 88.4\% for WRU with tract priors at the same coverage. Against the remaining tools Haiku stays ahead on common voters, by 6.5 points over WRU with tract priors and 3.8 points over rethnicity. Haiku's lead over pyethnicity in conditional accuracy comes from the voters it chooses to label.

\begin{figure}[htbp]
\centering
\includegraphics[width=\textwidth]{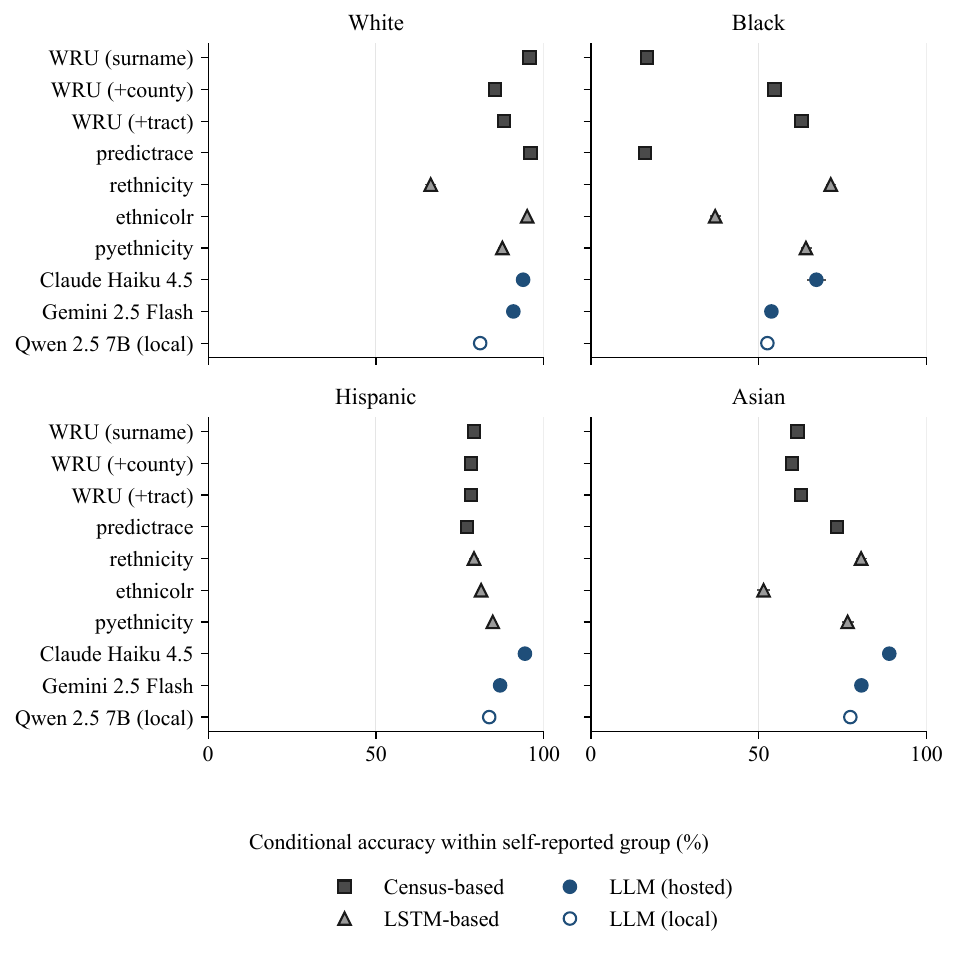}
\caption{Conditional accuracy within each self-reported group, Study 2 (5,000 White, 5,000 Black, 5,000 Hispanic, and 2,998 Asian voters, restricted to the voters each tool labels). Lines are 95\% bootstrap intervals, most of them narrower than the markers. Surname-only tools label most Black voters as White.}
\label{fig:by_race}
\end{figure}

The differences between tools are largest for Black voters (Figure~\ref{fig:by_race}). WRU with surnames only labels 16.7\% of Black voters correctly, and 82.0\% as White. Tract priors raise this to 62.8\%, pyethnicity reaches 64.1\%, rethnicity 71.5\%, and Haiku 67.2\% among the Black voters it labels. Because Haiku labels only a quarter of Black voters, it labels 16.8\% of all Black voters correctly, no more than surname-only WRU. Haiku also leads for Hispanic (94.4\%) and Asian (88.9\%) voters among those it labels. Haiku's Black F1 score, which balances precision and recall, is the highest of any tool among the voters it labels (0.79) but falls to 0.29 when unlabeled Black voters count as missed, above only predictrace and surname-only WRU (SI~\ref{app:pergroup}).

Haiku's coverage varies with the input and the wording of the request. Its coverage tracks the information in the name, falling from 53.8\% with full names to 28.8\% with surnames and 9.9\% with first names, while conditional accuracy stays near 90\% (SI~\ref{app:ablation}). But adding the voter's county of residence, which makes the name more informative, lowers Haiku's coverage to 31.1\%, and one wording of the request draws a refusal for 99\% of names (SI~\ref{app:prompts}). A request that looks more like a query about an identifiable person draws more refusals.

\section{Study 3: Cited Political Science Authors}
\label{sec:study3}

Studies 1 and 2 use benchmark data. Applied researchers work with samples that mix naming traditions, have no ground truth, and differ from the populations on which tools were validated. Study~3 therefore scores the tools on authors cited in political science journals.

We drew a stratified random sample of 300 authors from roughly 197,000 authors cited in 27 political science journals between 2020 and 2024, with ten strata that oversample informative cases (SI~\ref{app:strata}). A research assistant coded each author's gender and race from public information such as institutional profiles, without access to any prediction. A second coder independently coded 75 authors and agreed with the first on 86.7\% of codeable race classifications (Cohen's $\kappa = 0.81$) and 97.1\% of gender classifications ($\kappa = 0.94$). We use the first coder's labels as a human reference coding. They record perceived race and gender, not self-reports, and carry the uncertainty the reliability figures show. The coder placed 224 authors in one of the four race categories, could not determine race for 63, and coded 13 as other or mixed race, which lies outside the four categories. Coverage in Table~\ref{tab:study3_race} is the share of the 224 authors the tool labels, and accuracy is computed among authors whom both the tool and the coder classified.

\begin{table}[htbp]
\centering
\caption{Race and ethnicity against human reference coding, four categories (Study 3)}
\label{tab:study3_race}
\begin{threeparttable}
\small
\begin{tabular}{lccccc}
\toprule
 & \multicolumn{3}{c}{Unweighted (hard-case sample)} & \multicolumn{2}{c}{Weighted to frame} \\
\cmidrule(lr){2-4}\cmidrule(lr){5-6}
\textbf{Tool} & \textbf{Coverage} & \textbf{$N$} & \textbf{Accuracy [95\% CI]} & \textbf{Coverage} & \textbf{Accuracy} \\
\midrule
ethnicolr & 100.0 & 224 & 74.6 [68.3, 79.9] & 100.0 & 82.9 \\
pyethnicity & 100.0 & 224 & 71.0 [64.7, 76.8] & 100.0 & 78.3 \\
predictrace & 71.4 & 160 & 79.4 [72.5, 85.6] & 72.1 & 92.6 \\
rethnicity & 100.0 & 224 & 55.8 [49.6, 62.1] & 100.0 & 61.6 \\
WRU (surname) & 67.9 & 152 & 80.3 [73.7, 86.2] & 68.2 & 94.6 \\
\midrule
Gemini 2.5 Flash & 90.6 & 203 & 88.7 [84.7, 92.6] & 93.9 & 91.7 \\
Qwen 2.5 7B (local) & 77.2 & 173 & 83.8 [78.0, 89.0] & 81.1 & 91.4 \\
Claude Haiku 4.5 & 68.3 & 153 & 89.5 [85.0, 94.1] & 67.0 & 94.3 \\
\bottomrule
\end{tabular}
\begin{tablenotes}[flushleft]\small
\item Percent. Coverage is the share of the 224 authors with a four-category reference code whom the tool labels. Accuracy is computed on the $N$ authors whom both the tool and the coder classified. Unweighted figures describe the sample, which oversamples hard cases. Weighted figures use the design weights described in SI~\ref{app:study3full}. Intervals from 2,000 bootstrap replications.
\end{tablenotes}
\end{threeparttable}
\end{table}

The LLMs are the most accurate race classifiers in the sample (Table~\ref{tab:study3_race}). Haiku reaches 89.5\% and Gemini 88.7\%, against 80.3\% for WRU and 71.0\% for pyethnicity. The sample oversamples hard cases, however, including authors on whom WRU and an earlier LLM run disagreed. Weighted to the sampling frame, WRU (94.6\%) and Haiku (94.3\%) are about equally accurate, and each labels about two thirds of the authors. On the 101 authors both label, Haiku is correct on 97.5\% and WRU on 94.5\% after weighting (SI~\ref{app:study3full}). The LLM advantage is largest on hard cases, and much of the remaining LLM error lies across the Asian--White and Hispanic--White category boundaries (SI~\ref{app:study3full}). The LSTM tools lose the most relative to Study~2. Rethnicity falls from 73.7\% on North Carolina voters to 55.8\%, which fits a tool trained on U.S.\ voter data meeting an international name pool.

\begin{table}[htbp]
\centering
\caption{Gender against human reference coding (Study 3)}
\label{tab:study3_gender}
\begin{threeparttable}
\begin{tabular}{lccc}
\toprule
\textbf{Tool} & \textbf{$N$ codeable} & \textbf{Accuracy} & \textbf{95\% CI} \\
\midrule
Genderize.io & 229 & 97.4\% & [95.2, 99.1] \\
\texttt{gender} R package & 229 & 97.4\% & [95.2, 99.1] \\
\midrule
Claude Haiku 4.5 & 270 & 97.0\% & [94.8, 98.9] \\
Gemini 2.5 Flash & 258 & 98.1\% & [96.1, 99.6] \\
Qwen 2.5 7B (local) & 233 & 97.9\% & [95.7, 99.6] \\
\bottomrule
\end{tabular}
\begin{tablenotes}[flushleft]\small
\item Authors whom both the tool and the coder classified. An LLM answer of ``Unknown'' counts as no label. Intervals from 2,000 bootstrap replications.
\end{tablenotes}
\end{threeparttable}
\end{table}

For gender, every tool is correct on more than 97\% of the authors it labels (Table~\ref{tab:study3_gender}). The LLMs differ in coverage. Of the coded authors whom Genderize.io could not label, Haiku labels 45 and Gemini 38. In a mostly Western academic sample, then, the LLMs match the traditional tools on gender and reach authors those tools miss, and for race they hold their accuracy on the hard cases where traditional tools fail. The coder's labels are themselves inferences from public profiles, so Study~3 measures agreement with perceived race, not accuracy against self-reports.

\section{From Classification Accuracy to Downstream Bias}
\label{sec:error_propagation}

The three studies score tools by how often their labels are right, but a label that is usually right can still bias the group shares and differences estimated from it. This section measures that bias directly, using Study~2's self-reported race, every tool's prediction, and real outcomes for the same voters.

\subsection{Estimands and Empirical Confusion Matrices}
\label{sec:setup}

Let $R_i$ denote voter $i$'s self-reported race and $\hat{R}_i^{(t)}$ the label from tool $t$, which may be empty. The empirical confusion matrix $\mathbf{C}^{(t)}$ has entries $C^{(t)}_{ij}$ equal to the share of voters with $R = i$ who receive label $j$, with an extra column for voters who receive no label. We tabulate it directly from the predicted and true labels, so it involves no modeling assumption, and each row sums to one (SI~\ref{app:confusion}). Two entries anticipate the results. WRU with surnames only labels 82.0\% of Black voters as White. Haiku leaves 67.5\% of White, 75.0\% of Black, 18.2\% of Hispanic, and 29.4\% of Asian voters unlabeled.

We study three estimands with self-reported race as truth. The first is the Black share of the 19,998 sampled voters, $\pi_B = 0.25$. The second is the White--Black gap in Democratic registration, $\Delta_{\text{party}} = \Pr(\text{Dem} \mid \text{White}) - \Pr(\text{Dem} \mid \text{Black}) = -0.529$, where Democratic registration is 1 for Democrats and 0 for Republicans, unaffiliated voters, and members of minor parties. In the sample, 20.4\% of White and 73.3\% of Black voters are registered Democrats. The third is the White--Black gap in mean age, $\Delta_{\text{age}} = 4.29$ years (54.4 against 50.1). Party and birth year come from re-joining the sample to the voter file on name, race, ethnicity, gender, and county. Where the key matches several records we keep an outcome only if all of them agree. The gaps are therefore defined on the 19,418 voters with a party record and the 19,177 with a birth year, and every estimator, its oracle, and the decomposition below use the same voters. We do not redefine the estimand for the voters a pipeline labels, because that would hide selection bias in the definition. The 2,000 voters outside the four categories stay in the analysis, because a researcher cannot remove them before classifying and tools can label them White or Black. Dropping them changes no conclusion (SI~\ref{app:sensitivity}).

\subsection{Downstream Estimators on Real Outcomes}
\label{sec:estimators}

We compare four families of estimators and an oracle that uses true labels. \textit{Hard-label estimators} treat each tool's most likely category as the true one and compute the estimand among labeled voters, which is what most applied work does. \textit{Weighted estimators} use posterior probabilities, with $\hat{\pi}_j = N^{-1}\sum_i p_{ij}$ for shares and $\hat{\mu}_j = \sum_i p_{ij} Y_i / \sum_i p_{ij}$ for group means, where $p_{ij}$ is the probability a tool assigns to voter $i$ belonging to group $j$ \citep{chen2019fairness}. \textit{BIRDiE} fits the joint distribution of the outcome and race using BISG posteriors \citep{mccartan2024birdie}. In the main specification we use its categorical model for party registration (Democratic, Republican, unaffiliated, other) and its normal model for age, both pooled across counties and fitted by EM with the package's default weakly informative prior. Every fit converged. \textit{Mixed pipelines} use an LLM's label where it exists and a traditional tool's posterior where the LLM gave none. For inference we draw 500 bootstrap samples that resample voters within self-reported race, the sampling strata, and recompute every estimator and the true value in each sample. The intervals for bias are therefore paired, and the oracle's bias is zero in every sample. If our expectations hold, weighted estimates should understate the two gaps, BIRDiE should come close to them, and Haiku's hard labels should show bias from selection as well as from misclassification.

\begin{table}[!htbp]
\centering
\caption{Bias in downstream estimands, Study 2, with real outcomes from the voter file. The estimands are the Black share of the 19,998 sampled voters ($\pi_B = 0.25$), the White--Black gap in Democratic registration among the 19,418 voters with a party record ($\Delta_{\text{party}} = -0.529$, from 20.4\% of White and 73.3\% of Black voters registered as Democrats), and the White--Black gap in mean age among the 19,177 voters with a birth year ($\Delta_{\text{age}} = 4.29$ years, from 54.4 and 50.1 years). Brackets give 95\% paired bootstrap intervals for the bias (500 replicates).}
\label{tab:downstream_bias}
\footnotesize
\begin{tabular}{lccc}
\toprule
Estimator & Bias($\hat{\pi}_B$) & Bias($\hat{\Delta}_{\text{party}}$) & Bias($\hat{\Delta}_{\text{age}}$, years) \\
\midrule
\multicolumn{4}{l}{\textit{Hard-label (argmax) estimators}} \\
\quad WRU surname & \shortstack{$-0.195$\\{}[$-0.197$, $-0.191$]} & \shortstack{$+0.334$\\{}[$+0.303$, $+0.363$]} & \shortstack{$-3.07$\\{}[$-4.36$, $-1.87$]} \\
\quad WRU +county & \shortstack{$-0.031$\\{}[$-0.036$, $-0.025$]} & \shortstack{$+0.298$\\{}[$+0.276$, $+0.317$]} & \shortstack{$-2.62$\\{}[$-3.45$, $-1.86$]} \\
\quad WRU +tract & \shortstack{$-0.023$\\{}[$-0.028$, $-0.018$]} & \shortstack{$+0.240$\\{}[$+0.222$, $+0.259$]} & \shortstack{$-2.69$\\{}[$-3.46$, $-1.97$]} \\
\quad pyethnicity & \shortstack{$-0.028$\\{}[$-0.032$, $-0.024$]} & \shortstack{$+0.257$\\{}[$+0.240$, $+0.275$]} & \shortstack{$-0.52$\\{}[$-1.29$, $+0.29$]} \\
\quad Claude Haiku 4.5 & \shortstack{$-0.153$\\{}[$-0.158$, $-0.148$]} & \shortstack{$+0.191$\\{}[$+0.154$, $+0.222$]} & \shortstack{$+6.04$\\{}[$+4.67$, $+7.44$]} \\
\quad Gemini 2.5 Flash & \shortstack{$-0.089$\\{}[$-0.093$, $-0.085$]} & \shortstack{$+0.271$\\{}[$+0.251$, $+0.290$]} & \shortstack{$+1.42$\\{}[$+0.58$, $+2.29$]} \\
\quad Qwen 2.5 7B (local) & \shortstack{$-0.046$\\{}[$-0.051$, $-0.041$]} & \shortstack{$+0.344$\\{}[$+0.322$, $+0.363$]} & \shortstack{$+0.53$\\{}[$-0.35$, $+1.52$]} \\
\quad Hybrid Haiku$\rightarrow$pyeth. & \shortstack{$-0.038$\\{}[$-0.042$, $-0.034$]} & \shortstack{$+0.262$\\{}[$+0.245$, $+0.280$]} & \shortstack{$-0.11$\\{}[$-0.88$, $+0.72$]} \\
\addlinespace
\multicolumn{4}{l}{\textit{Posterior-weighting estimators}} \\
\quad WRU surname & \shortstack{$-0.099$\\{}[$-0.101$, $-0.097$]} & \shortstack{$+0.426$\\{}[$+0.410$, $+0.440$]} & \shortstack{$-3.82$\\{}[$-4.51$, $-3.07$]} \\
\quad WRU +county & \shortstack{$-0.030$\\{}[$-0.033$, $-0.027$]} & \shortstack{$+0.367$\\{}[$+0.351$, $+0.381$]} & \shortstack{$-3.43$\\{}[$-4.10$, $-2.73$]} \\
\quad BISG tract chain & \shortstack{$-0.025$\\{}[$-0.028$, $-0.022$]} & \shortstack{$+0.307$\\{}[$+0.293$, $+0.320$]} & \shortstack{$-3.32$\\{}[$-3.97$, $-2.67$]} \\
\quad pyethnicity & \shortstack{$-0.003$\\{}[$-0.006$, $-0.001$]} & \shortstack{$+0.323$\\{}[$+0.308$, $+0.337$]} & \shortstack{$-1.03$\\{}[$-1.68$, $-0.34$]} \\
\addlinespace
\multicolumn{4}{l}{\textit{Mixed LLM + posterior pipelines}} \\
\quad Haiku + pyeth.\ posterior & \shortstack{$-0.019$\\{}[$-0.022$, $-0.016$]} & \shortstack{$+0.316$\\{}[$+0.301$, $+0.330$]} & \shortstack{$-0.47$\\{}[$-1.15$, $+0.23$]} \\
\quad Haiku + BISG posterior & \shortstack{$-0.037$\\{}[$-0.040$, $-0.034$]} & \shortstack{$+0.283$\\{}[$+0.267$, $+0.297$]} & \shortstack{$-1.38$\\{}[$-2.02$, $-0.73$]} \\
\addlinespace
\multicolumn{4}{l}{\textit{BIRDiE estimators}} \\
\quad BIRDiE, WRU surname & \shortstack{$-0.096$\\{}[$-0.098$, $-0.094$]} & \shortstack{$-0.029$\\{}[$-0.069$, $+0.007$]} & \shortstack{$+1.23$\\{}[$-1.70$, $+3.47$]} \\
\quad BIRDiE, WRU +county & \shortstack{$-0.027$\\{}[$-0.030$, $-0.024$]} & \shortstack{$-0.012$\\{}[$-0.041$, $+0.013$]} & \shortstack{$+0.60$\\{}[$-0.90$, $+1.95$]} \\
\quad BIRDiE, BISG tract chain & \shortstack{$-0.022$\\{}[$-0.025$, $-0.019$]} & \shortstack{$-0.020$\\{}[$-0.042$, $+0.001$]} & \shortstack{$-0.53$\\{}[$-1.63$, $+0.51$]} \\
\addlinespace
\multicolumn{4}{l}{\textit{Reference}} \\
\quad Oracle (true labels) & 0 (by construction) & 0 (by construction) & 0 (by construction) \\
\addlinespace
\bottomrule
\end{tabular}
\begin{minipage}{0.97\textwidth}\vspace{4pt}\footnotesize
BIRDiE's Black share comes from the party model. The BISG tract chain uses tract priors for the 79\% of voters whose address geocoded to a tract and county priors otherwise.
\end{minipage}
\end{table}

\begin{figure}[htbp]
\centering
\includegraphics[width=\textwidth]{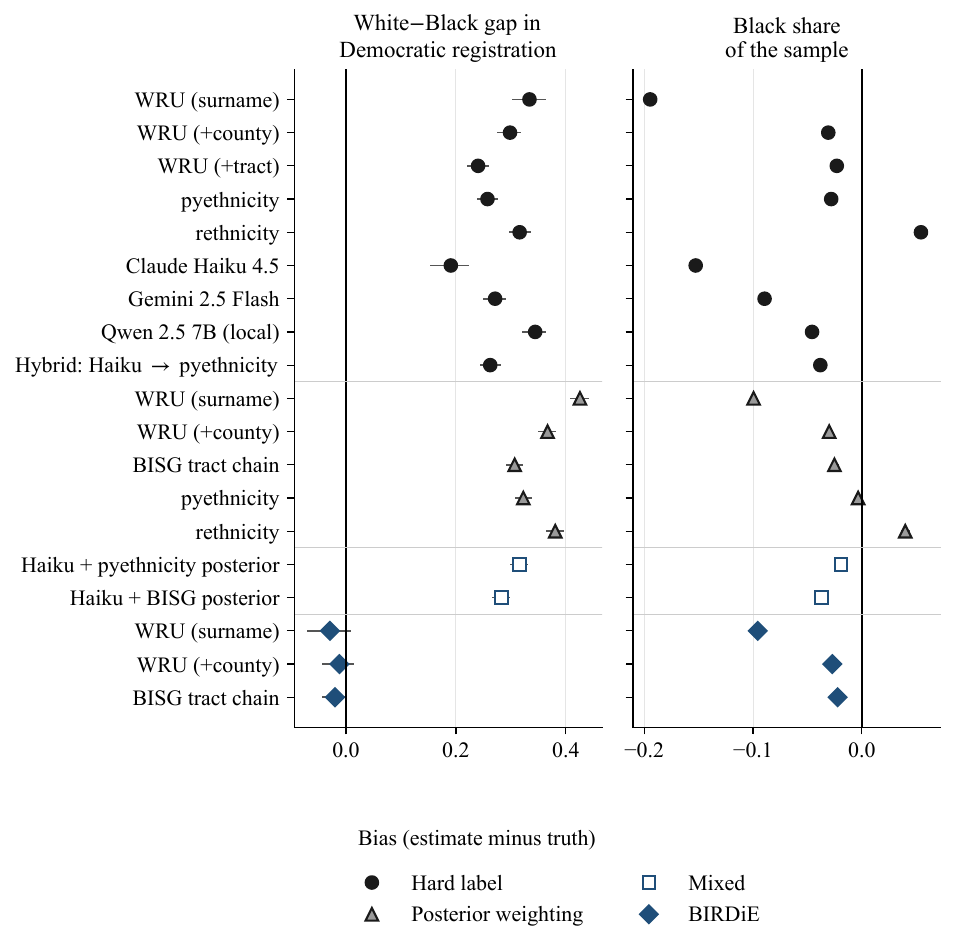}
\caption{Bias of each estimator, Study 2. Left, the White--Black gap in Democratic registration (true gap $-0.529$, $N = 19{,}418$ voters with party recorded). Right, the Black share of the sample (true share 0.25, $N = 19{,}998$). Lines are 95\% paired bootstrap intervals for the bias. Only the BIRDiE estimates come close to the true registration gap.}
\label{fig:downstream_bias}
\end{figure}

Table~\ref{tab:downstream_bias} and Figure~\ref{fig:downstream_bias} report the results. For the Black share, calibration and coverage decide the ranking. Weighting by pyethnicity's posteriors misses the true share by 0.3 percentage points, and weighting by BISG posteriors with county or tract priors comes within 3 points, while surname-only posteriors miss by 10. Selective hard labels fail. Haiku's labels imply a Black share of 9.7\% against a true 25\%, and surname-only WRU implies 5.5\% because it labels most Black voters White.

For the registration gap, only BIRDiE comes close. On county-level BISG posteriors it estimates the gap at $-0.541$ against a true $-0.529$, a bias of $-0.012$ (95\% interval $-0.041$ to $+0.013$), and the surname-only and tract versions miss by 0.029 and 0.020. All three bias intervals contain zero. Every other estimator understates the gap, hard labels by 36\% to 65\% of its size and posterior weighting by 58\% to 81\%. Haiku is the least biased hard-label estimator ($+0.191$), consistent with its high conditional accuracy, and it misses by more than each BIRDiE estimate in every bootstrap sample. The weighted estimators attenuate most, as \citet{chen2019fairness} lead one to expect when a name's probability is spread across several groups.

BIRDiE's result depends on how it treats geography, and in this benchmark the dependence tracks how informative the posteriors are (Table~\ref{tab:birdie_variants} in the SI). County fixed or random effects ask BIRDiE to estimate a race-by-party table for each county. Counties in the sample have a median of 78.5 voters, and with diffuse posteriors the model cannot separate White from Black voters within a county and pulls their rates together. Surname-only posteriors give Black voters a mean probability of 0.31 of being Black, and their county models understate the gap by 0.092 and 0.120. Tract priors raise that probability to 0.55, and every tract specification stays within 0.023 of the true gap. Restricting the analysis to voters geocoded to a tract gives similar results. The disagreement between specifications tracks the error. Across the three posteriors, the range of the pooled, fixed-effect, and random-effect estimates is 0.149, 0.082, and 0.043, and the largest absolute bias is 0.120, 0.070, and 0.023. The range needs no ground truth and can serve as a sensitivity check, though agreement across specifications does not establish that the estimates are accurate. Within self-reported race and county, we detect no association between either outcome and how Black a surname looks ($p > 0.19$). This limited test does not establish BIRDiE's independence assumption. This robustness holds for the party gap. For the age gap, county fixed effects move the tract estimate's bias from $-0.53$ to $-2.47$ years.

The age gap shows the cost of informative missingness. Haiku's hard labels overstate it by 6.0 years (95\% interval 4.7 to 7.4), the largest absolute bias in the age gap among the estimators compared. Haiku's missing labels are related to age within racial groups. Its ``Unknown'' answers fall on older Black voters (mean age 55.0 against 50.1 for all Black voters), while its explicit refusals fall on younger ones (SI~\ref{app:confusion}), and no reweighting of the labeled voters by race can remove a bias that operates within races. Filling Haiku's missing labels with pyethnicity's labels brings the age bias to $-0.11$ years ($-0.88$ to $+0.72$), because the fallback returns the unlabeled voters to the estimand. It does not remove misclassification, and the same hybrid understates the registration gap by 0.262, more than Haiku alone. Gemini's and Qwen's hard labels also overstate the age gap, by 1.4 and 0.5 years, while every traditional tool's hard labels understate it, so hard-label bias has no common direction. BIRDiE misses the age gap by between $-0.53$ and $+1.23$ years depending on the posterior, with intervals that contain zero, and a paired comparison cannot rank it against the hybrid.

Reweighting the sample to the race composition of the North Carolina voter file changes the magnitudes but not the conclusions (SI~\ref{app:popdownstream}).

\subsection{Decomposing the Bias and Testing the Complementarity Claim}
\label{sec:decision}

\begin{figure}[htbp]
\centering
\includegraphics[width=\textwidth]{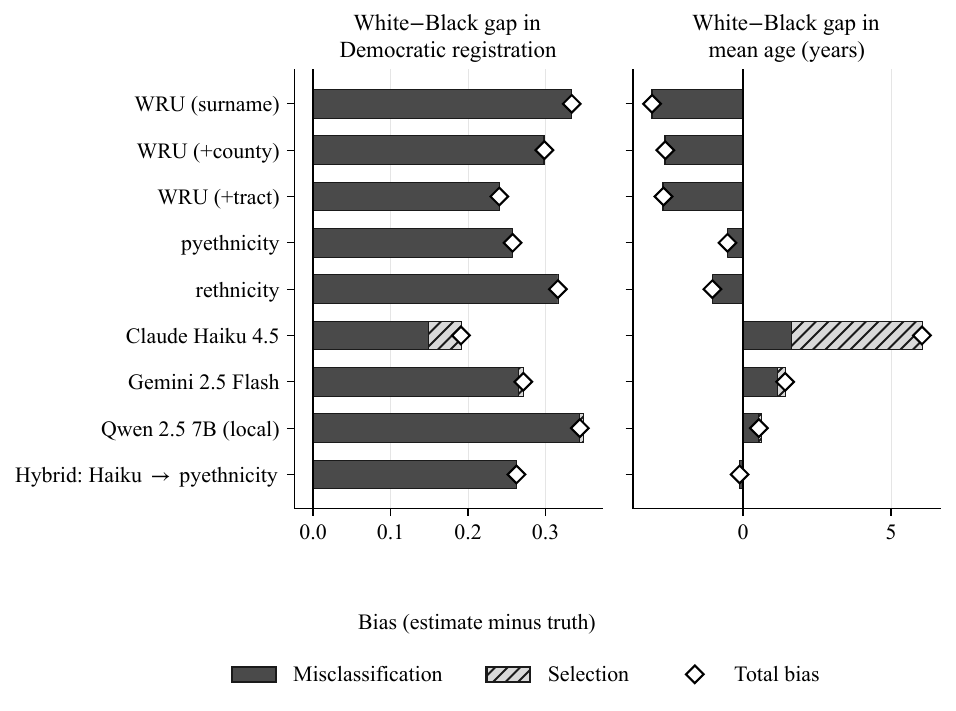}
\caption{Hard-label bias split into misclassification and selection by Equation~\ref{eq:decomposition}, Study 2 ($N = 19{,}418$ with party recorded and $N = 19{,}177$ with a birth year). Diamonds mark total bias. Pipelines that label every voter have no selection component. Most of Haiku's bias in the age gap comes from selection.}
\label{fig:bias_decomposition}
\end{figure}

Because we observe true race for every voter, the bias of each hard-label estimator splits exactly into a misclassification part and a selection part. Let $S_t$ be the set of voters tool $t$ labels and $\hat{\Delta}_{\text{oracle}}(S_t)$ the estimate from true labels within $S_t$. Then
\begin{equation}
\underbrace{\hat{\Delta}_t - \Delta}_{\text{total bias}}
= \underbrace{\bigl[\hat{\Delta}_t - \hat{\Delta}_{\text{oracle}}(S_t)\bigr]}_{\text{misclassification}}
+ \underbrace{\bigl[\hat{\Delta}_{\text{oracle}}(S_t) - \Delta\bigr]}_{\text{selection}},
\label{eq:decomposition}
\end{equation}
an identity with no assumptions or free parameters. The selection part is zero for any pipeline that labels everyone. Figure~\ref{fig:bias_decomposition} shows both parts for each tool. For the traditional tools, bias comes almost entirely from misclassification. Haiku fails mostly through selection in the age gap, where 4.4 of its 6.0 years of bias come from which voters it left unlabeled, almost all of it through its ``Unknown'' answers (SI~\ref{app:confusion}). Gemini and Qwen label nearly every voter, so their selection components are small. Haiku's case is the regime the reject-option literature warns about. A correction that applies a single confusion matrix, unrelated to the outcome, cannot reach it, because the selection operates within racial groups. Models that let missingness depend on the outcome could address it in principle, but they need information about the unlabeled voters.

The complementary workflow pays off only if LLMs are accurate on the names traditional tools get wrong. Table~\ref{tab:complementarity} conditions each LLM's accuracy on whether a traditional tool labels the name correctly. On the 11,418 names that surname-only WRU gets right, Haiku labels 62.0\% and is correct on 98.3\%. On the 6,570 names WRU gets wrong, Haiku labels 30.5\% and is correct on 57.5\%. Gemini and Qwen, which rarely refuse, show the same drop, to 45.3\% and 45.0\% on WRU's errors. On pyethnicity's errors the LLMs are right on 11.4\% to 21.3\% of the names they label. The tools often fail on the same names, and refusal steers Haiku toward the easy ones. None of the tested fallback pipelines exceeds pyethnicity's total accuracy in this sample (Table~\ref{tab:study2_overall_v2}).

\begin{table}[!htbp]
\centering
\caption{LLM accuracy conditional on whether a traditional tool classifies the name correctly, Study 2 four-category sample ($N = 17{,}998$). Coverage is the share of names in the row that the LLM labels. Conditional accuracy is the share correct among labeled names.}
\label{tab:complementarity}
\small
\begin{tabular}{lccccc}
\toprule
 & & \multicolumn{2}{c}{Claude Haiku 4.5} & Gemini 2.5 Flash & Qwen 2.5 7B \\
\cmidrule(lr){3-4}
Names the traditional tool gets & $N$ & Coverage & Cond.\ acc. & Cond.\ acc. & Cond.\ acc. \\
\midrule
WRU (surname) correct & 11,418 & 62.0 & 98.3 & 95.6 & 89.2 \\
WRU (surname) wrong &  6,570 & 30.5 & 57.5 & 45.3 & 45.0 \\
WRU (+tract) correct & 13,291 & 56.5 & 96.4 & 89.5 & 83.6 \\
WRU (+tract) wrong &  4,641 & 33.4 & 54.9 & 45.4 & 43.5 \\
pyethnicity correct & 14,121 & 57.9 & 97.9 & 93.5 & 87.5 \\
pyethnicity wrong &  3,877 & 23.3 & 11.4 & 16.9 & 21.3 \\
\bottomrule
\end{tabular}
\end{table}

\subsection{Combining LLM Answers with BISG Posteriors}
\label{sec:combination}

A researcher can also combine LLMs with probabilistic tools. We treat each LLM's answer as evidence about race and update the tract BISG posterior by Bayes' rule, $\Pr(R \mid S, G, L) \propto \Pr(R \mid S, G) \Pr(L \mid R)$, where $S$ is the surname, $G$ the geography, and $L$ the LLM's answer or its type of missing output. We estimate $\Pr(L \mid R)$ on one half of the sample and apply it to the other, so that no voter's self-report enters the voter's own posterior (SI~\ref{app:combination}). Every LLM's answers raise the accuracy of the most likely category, from 74.1\% for BISG alone to between 77.9\% and 79.4\%. Combined with Gemini's answers, the posteriors are 1.0 point more accurate than pyethnicity, the most accurate single tool (95\% interval 0.4 to 1.5). Haiku's answers help although half of them are missing, because whether and how Haiku declines carries information about race. The weighted estimator gains as well. Its bias for the age gap shrinks from $-3.32$ years to between $-1.36$ and $-1.93$, and its bias for the registration gap from 0.307 to between 0.261 and 0.280, which still leaves half of the gap missing.

The combination provides no demonstrated gain in BIRDiE accuracy. On BISG posteriors alone its bias intervals contain zero for both gaps, and on every combined posterior they exclude it, with biases of $+0.034$ to $+0.054$ for the registration gap and $+0.96$ to $+2.12$ years for the age gap. BISG alone errs slightly in the other direction, and the paired differences in absolute bias are not statistically distinguishable from zero (SI~\ref{app:combination}). The update treats the LLM's answer as independent of the surname and geography given race, even though the LLM reads the same surname. It therefore counts surname evidence twice, and the resulting probabilities are overconfident. First names may also carry birth-cohort information that violates BIRDiE's outcome--name independence assumption, but this design does not separate that explanation from miscalibration. Held-out calibration improves prediction without reducing BIRDiE's bias relative to original BISG (SI~\ref{app:combination}).

\section{Discussion}
\label{sec:discussion}

We asked whether the accuracy of LLM name classifiers carries over to the estimates researchers report, and found that it does so only in part. Haiku's apparent advantage over pyethnicity disappears on common voters, yet its missing labels produce selection bias that a single confusion-matrix correction cannot remove. For the registration gap, the choice of estimator matters more than the choice of classifier. The LLMs still add information. Combined with BISG, Gemini's answers give the most accurate individual labels in the study, and for Chinese names Gemini's majority agreement stays at 87.7\% on the names Genderize.io cannot label.

Conditional accuracy, the metric on which LLMs look strongest, does not describe how well a classifier serves an estimate. The classifier with the highest conditional accuracy in our benchmark produced the largest absolute bias in the age gap. A classifier that abstains on hard cases trades coverage for accuracy, with the trade only remaining harmless when abstention is unrelated to the outcome \citep{chow1970optimum, elyaniv2010foundations}, which is a condition Haiku's missing labels do not meet. The selection runs mainly through Haiku's ``Unknown'' answers, but the output categories do not establish whether uncertainty or safety training explains it. This pattern is consistent with distinctive first names being more common in younger cohorts, although our design cannot establish the mechanism. Applied work should therefore account for selection when handling missing LLM labels. For the LLMs that rarely refuse in this benchmark, misclassification accounts for most of the bias.

The results make the intended estimand central to tool choice. To estimate racial gaps in the United States, BIRDiE turns decade-old BISG posteriors into close estimates of a gap for a discrete outcome, and comparing geographic specifications reveals sensitivity to modeling choices. The tested LLM--BISG update is overconfident, so its gains in individual classification do not establish its suitability for BIRDiE. For individual classification, the LLMs earn a place. Their answers sharpen BISG posteriors, and a voter Haiku labels Black is Black 95.8\% of the time, the highest precision of any tool (SI~\ref{app:pergroup}). The combination needs error rates estimated on people with known race, and so a validation sample. LLMs also extend coverage to populations where no validated estimator of downstream disparities exists, such as gender in non-Western naming systems and race for authors outside the United States. Genderize.io returns probabilities, but a probability for each name is not an estimator that corrects downstream bias. In these populations the LLMs label names traditional tools miss, but whether their labels yield unbiased estimates is untested, and a fallback restores unlabeled cases without correcting its own errors. Table~\ref{tab:checklist} turns these results into a checklist.

\begin{table}[htbp]
\centering
\caption{Choosing a classification pipeline by goal and population}
\label{tab:checklist}
\footnotesize
\begin{tabularx}{\textwidth}{>{\raggedright\arraybackslash}p{3.2cm}>{\raggedright\arraybackslash}X>{\raggedright\arraybackslash}X}
\toprule
\textbf{Goal} & \textbf{U.S.\ race with names and addresses} & \textbf{No validated downstream estimator (gender, non-U.S.\ names)} \\
\midrule
Group shares & Validate full-coverage posteriors for the target population before using them to estimate shares. Account for selection if a tool leaves cases unlabeled. & Do not assume the labeled subset represents the full sample. A fallback restores coverage but not correct labels, so report coverage by group and validate the shares on a subsample with known identity. \\
\addlinespace
Group differences, discrete outcome & Evaluate BIRDiE on BISG posteriors and report sensitivity to geographic specifications. Validate calibration and estimator assumptions before adding LLM information. Weighted estimates attenuated this benchmark's registration gap. & Report estimates with and without a fallback, and establish the direction of bias on a validation subsample. \\
\addlinespace
Group differences, continuous outcome & Compare BIRDiE with a full-coverage pipeline. Hard-label bias can inflate or shrink a gap. Check the direction of bias in weighted estimates on a validation sample. & As above, and check on the validation subsample whether missing labels are related to the outcome within groups. \\
\addlinespace
Individual labels & Evaluate combinations of BISG posteriors and LLM answers, missing outputs included, using error rates estimated on voters with known race. Check calibration if probabilities will be used. & Use LLM labels where traditional tools give none. Validate their accuracy and report coverage by group, because the labeled names are a selected subset. \\
\addlinespace
Any goal & \multicolumn{2}{>{\raggedright\arraybackslash}p{11.5cm}}{Report coverage, conditional accuracy, and total accuracy, and compare tools on common cases and at matched coverage. Document who goes uncounted \citep{lee2025soda}. Archive tool versions, prompts, and raw outputs, and prefer a locally versioned model when results must be reproducible over time \citep{barrie2025replication}.} \\
\bottomrule
\end{tabularx}
\end{table}

The downstream analysis rests on one state and two outcomes, so its magnitudes are an existence proof of selection bias, not general estimates. The LLMs may have seen the public North Carolina voter file during training. In a temporal holdout, voters who registered after each model's training cutoff are labeled about as accurately as earlier registrants (for Haiku's 675 later registrants a difference of $-1.1$ points, 95\% interval $-5.0$ to $+2.7$, SI~\ref{app:temporal}). Neither this test nor Study~3, whose authors were never compiled into a public dataset, rules out exposure to name and race associations in general \citep{argyle2024misclassification}.

Hosted models change, so our hosted results describe the model versions we queried. The local model reproduces exactly, but it trails the hosted models on conditional accuracy.

Study~3 is small, with 153 authors labeled by both Haiku and the coder for race, and its reference codes record one coder's perception of race. Our gender classification scheme does not represent nonbinary identities \citep{lockhart2023namebased, keyes2018misgendering}, and every tool assigns categories to people who did not choose them \citep{gautam2024stop}. Inferring identity from names requires justification, and the justification weakens when the inferred categories are fed into estimators that distort them.

The question for name-based demographic research is how a classifier's errors and refusals move the estimate a study reports, and accuracy rankings do not answer it. In our voter benchmark, BIRDiE on BISG posteriors comes closest to the registration gap, while LLM answers improve individual classification when combined with BISG. Where no validated estimator exists, LLMs label names that traditional tools miss, but their missing labels must be accounted for and their labels validated against the estimand before they can be trusted.

\section*{Statements and Declarations}

\subsection*{Ethical considerations}
The study analyzes publicly available records (a state voter registration file, a published name-frequency dataset, and public information about published academic authors) and involved no interaction with human participants.

\subsection*{Consent to participate}
Not applicable.

\subsection*{Consent for publication}
Not applicable.

\subsection*{Declaration of conflicting interest}
The author(s) declared no potential conflicts of interest with respect to the research, authorship, and/or publication of this article.

\subsection*{Funding statement}
The author(s) received no financial support for the research, authorship, and/or publication of this article.

\subsection*{Data availability}
\ifdefined\preprintversion
Analysis code and replication materials will be made public through Harvard Dataverse upon acceptance.
\else
Analysis code and replication materials are available through an anonymous reviewer link at \href{https://dataverse.harvard.edu/previewurl.xhtml?token=23de2e10-d9c5-4d66-8c43-754b637978c1}{\url{https://dataverse.harvard.edu/previewurl.xhtml?token=23de2e10-d9c5-4d66-8c43-754b637978c1}} and will be made public through Harvard Dataverse upon acceptance.
\fi
The package includes Study~1 name-frequency records and Study~2 analytical records with direct identifiers removed. Archived classifier outputs allow reviewers to refit the downstream analyses, including the LLM--BISG combinations and calibration check. Study~3 is shared only as aggregate results to protect the scholars whose perceived identities were coded. Source-data instructions and checksums document reconstruction of Studies~1 and~2; rerunning hosted classifiers may produce different outputs. The North Carolina voter file and Chinese name dataset are available from their original sources.

\subsection*{Use of generative AI}
Three large language models (Claude Haiku 4.5, Gemini 2.5 Flash, and Qwen 2.5 7B Instruct) are the objects of study and were used as classifiers in all three studies. The authors designed the study, analyzed the data, and take full responsibility for the content. Sonnet 5 (Anthropic) was used for language editing during manuscript preparation, Claude Code Opus 5.5 was used for preparing the replication material.

\bibliography{biblio}

\ifdefined\preprintversion
\preprintappendix
\fi
\end{document}